\documentclass[amsmath,amssym,amsfonts,pre,twocolumn, showkeys]{revtex4}

\usepackage{graphicx}
\usepackage{epstopdf}

\newcommand{\xv}{{\bf x}}
\newcommand{\Xv}{{\bf X}}

\newcommand{\uv}{{\bf u}}

\newcommand{\tv}{{\bf t}}

\newcommand{\bD}{{\bf \Delta}}
\newcommand{\grad}{{\bf \nabla}}

\begin{document}

\title{How geometric frustration shapes twisted fibers, inside and out: \\ Competing morphologies of chiral filament assembly}

\author{Douglas M. Hall}  
\author{Gregory M. Grason}
\affiliation{Department of Polymer Science and Engineering, University of Massachusetts Amherst, USA} 

\begin{abstract} 

Chirality frustrates and shapes the assembly of flexible filaments in rope-like, twisted bundles and fibers by introducing gradients of both filament shape (i.e. curvature) and packing throughout the structure.   Previous models of chiral filament bundle formation have shown that this frustration gives rise to several distinct morphological responses, including self-limiting bundle widths, anisotropic domain (tape-like) formation and topological defects in the lateral inter-filament order.  In this paper, we employ a combination of continuum elasticity theory and discrete filament bundle simulations to explore how these distinct morphological responses compete in the broader phase diagram of chiral filament  assembly.  We show that the most generic model of bundle formation exhibits at least four classes of equilibrium structure -- finite-width, twisted bundles with isotropic and anisotropic shapes, with and without topological defects, as well as bulk phases of untwisted, columnar assembly (i.e. ``frustration escape"). These competing equilibrium morphologies are selected by only a relatively small number of parameters describing filament assembly:  bundle surface energy, preferred chiral twist and stiffness of chiral filament interactions, and mechanical stiffness of filaments and their lateral interactions.  Discrete filament bundle simulations test and verify continuum theory predictions for dependence of bundle structure (shape, size and packing defects of 2D cross section) on these key parameters.
\end{abstract}

\keywords{Chirality; Geometric Frustration; Self Assembly; Filaments; Fibers } 

\maketitle

\section{Introduction}

Assemblies of chiral building blocks often inherit the lack of mirror symmetry at the mesoscale from the microscopic scale of those subunits \cite{Harris1999}.  This effect is well known in the context of chiral mesophases of liquid crystals \cite{Straley1976, Goodby1991}, in which the lack of inversion symmetry of rod-like molecules leads to self-organized, twisted textures. One such state is the cholesteric phase, where the molecular axis adopts a handed twist throughout the sample.  In a second class of systems, membranes formed by chiral molecules (e.g. surfactants), the local 2D packing of chiral elements ultimately leads to mesoscopically handed morphologies, including helicoidal and spiral ribbons \cite{Selinger2001, Ghafouri2005, Armon2014}.  In biological assemblies, where protein building blocks are universally chiral at the molecular scale, the effect of chirality to template morphology at a super-molecular length scale is well known \cite{neville1993biology, Bouligand2008a, Sharma2009}.  

\begin{figure*}
\centerline{
\includegraphics[width=0.9\linewidth]{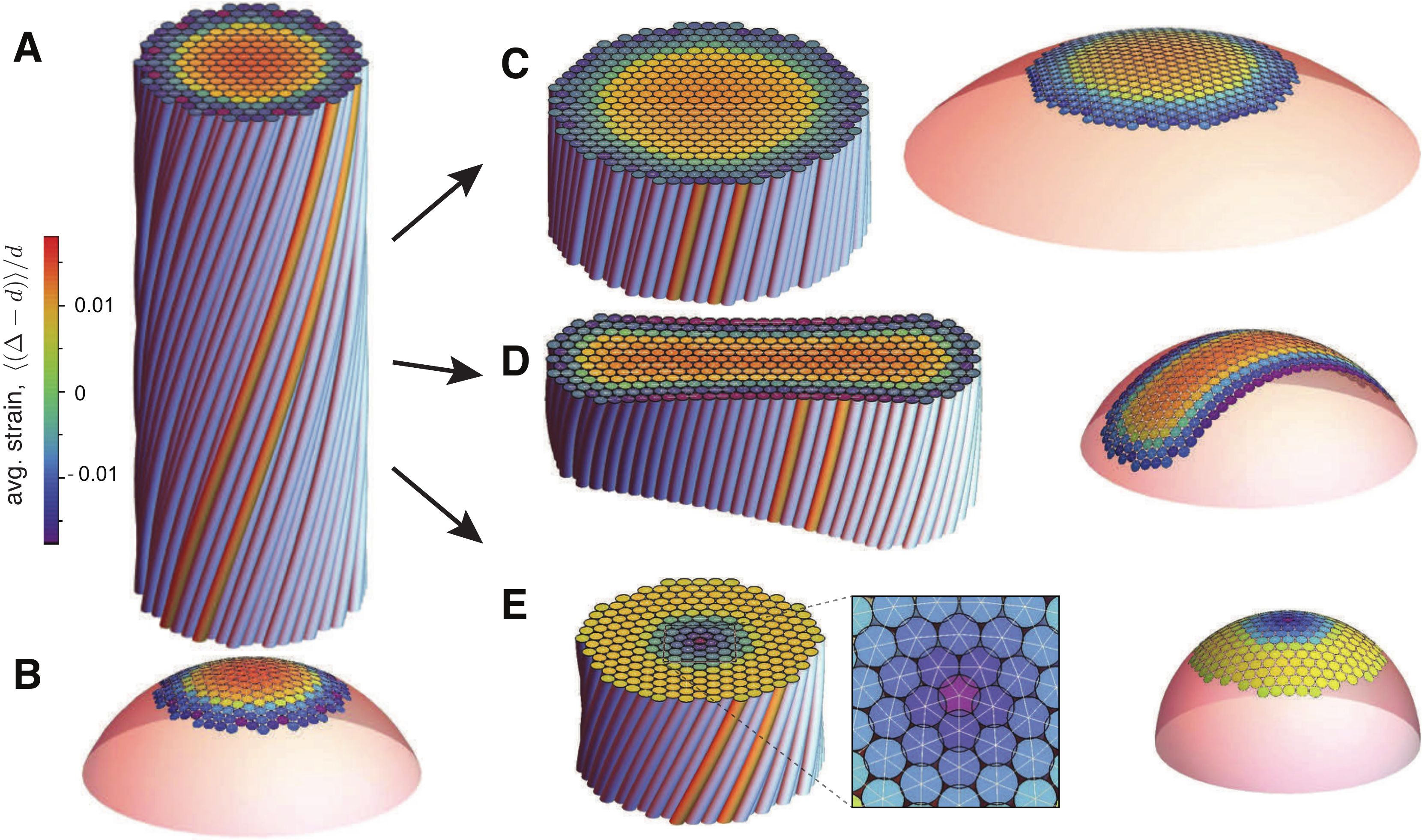}
}
\caption{\label{Figure 1}
Competing morphologies of self-assembled chiral filaments. Compression at the perimeter is due to geometric frustration of twisted textures. We show equivalent bundle cross section (A) and curved crystal structure (B) using the previously derived mapping. At the top, elements are shaded according to local strain of inter filament distance $\Delta$, relative to ideal separation $d$. In (C), unwinding preferred chiral twist (or equivalently decreasing spherical curvature) allows a larger assembly with reduced compression. In (D), excess free surface of anistropic, tape-like bundles leads to compression that grows only with thinner cross-section dimension. In (E), the introduction of a 5-fold defect in the inter-filament order relaxes hoop compression.
}
\end{figure*}

In this paper, we consider the influence of chirality to shape rope-like assemblies of 1D filamentous elements, a common materials architecture.  Chiral bundles and fibers form in a diverse range of systems, from supramolecular polymers, organogels and columnar fibers \cite{Brunsveld2001, Chakrabarti2009, Douglas2009, Prybytak2012, Wang2013}, to protein fibers, both functional (collagen \cite{Prockop1998, wess2008collagen}, cellulose \cite{Fratzl2003}, fibrin \cite{Weisel1987}) and pathological (amyloids \cite{Rubin2008}, sickle hemoglobin \cite{Makowski1986}).  While these systems are clearly diverse in terms of both their structure and interactions at the molecular scale, they share a common geometrical template, with filaments or columnar molecular stacks winding helically along the long axis of the bundle or fiber (see Fig. \ref{Figure 1}A).  This basic geometry leads to generic, yet non-trivial, consequences for the resulting structures and thermodynamics of assembly, owing to the geometrical frustration of long-range order by chiral textures in self-organized bundles.  

Geometric frustration refers to the incompatibility of preferred patterns of local order and global constraints, which prevents the propagation of distortion-free order in assemblies \cite{Kleman1989, Sadoc2006, Grason2016}.  In rope-like twisted bundles, frustration arises from two geometric mechanisms.  The first,  {\it orientational frustration},  results from the tendency of chirality to promote non-parallel textures of filament backbone orientation throughout the assembly \cite{Grason2009}.   Lateral gradients of filament bending introduced by chiral ordering give rise to important size-dependent assembly costs that can establish the equilibrium (self-limiting) finite size of bundles \cite{Turner2003a, Grason2007a, Yang2010, Brown2014}.  The second mechanism is {\it metric frustration}, where the non-parallel, twisted textures of filaments favored by chirality make uniform {\it inter-filament} spacing impossible \cite{Grason2015}.  It has been shown that this metric frustration in twisted bundles can be mapped onto the metric frustration of packing elements on a positively-curved 2D surface \cite{Bruss2012a, Bruss2013}.  Specifically, the distance of closest approach between filaments in a twisted bundle corresponds one-to-one to the geodesic distance between equivalent points on the ``dual surface", whose curvature radius is proportional to the helical pitch of bundle twist (see Fig. \ref{Figure 1}B).  This ``dual problem" has received broad interest in diverse contexts ranging from the long-standing Thomson problem \cite{Bowick2009} to the structure of particle-assemblies at liquid droplet interfaces \cite{Manoharan2015}.  Previous studies have shown that metric frustration in both bundles  and curved crystals leads to intrinsic gradients of stress in the assembly that in turn may trigger a rich set of morphological responses \cite{Grason2016}, including self-limiting domains \cite{Turner2003a, Grason2007a}, defective bulk order \cite{Seung1988, Bausch2003a, Bowick2009, Grason2010a, Grason2012} and anisotropic surface shapes \cite{Schneider2005, Morozov2010, Meng2014, Hall2016}.

In this paper, we explore the principles that select among the multiple competing morphologies stabilized by geometric frustration in chiral filament bundles.  These distinct morphological responses are illustrated in Fig. \ref{Figure 1}, associated with bundle geometry {\it along the bundle} (Fig. \ref{Figure 1}C), bundle surface geometry (Fig. \ref{Figure 1}D) and symmetries of inter-filament order in the bulk (Fig. \ref{Figure 1}E).  While chiral filament interactions select a preferred amount of bundle twist, the cost of both intra- and inter-filament strains may be relaxed via an untwisting of skewed filament packing (Fig. \ref{Figure 1}C), which corresponds to a geometric  ``flattening" of the dual metric surface as lateral bundle area grows and allows bundles to reach a larger diameter than possible for fixed twist.  Indeed, when chiral interactions are sufficiently ``soft', bundle assemblies may ``escape" frustration by untwisting completely, and consequently grow to unbounded domains.  Alternatively, a second mechanism to relax frustration cost is possible if surface energies (associated with fewer cohesive bonds at the bundle exterior) are sufficiently small compared to the cost to deform inter-filament bonds, so that the bundle cross section becomes highly anisotropic as shown in the tape-like morphology in Fig. \ref{Figure 1}D.  Analogous to the elastic instability for growing crystalline domains on spherical surfaces described below, this morphology is driven by a tendency to lower the overall growth of inter-filament strains with increasing filament number in the bundle.  Finally, Fig. \ref{Figure 1}E shows that for sufficiently high overall twist in the assembly, it becomes advantageous to adopt (5-fold) disclination defects in the 2D bundle cross section to lower the cost of geometric frustration of inter-filament spacing, a phenomenon well-studied for the assembly of crystalline arrays of positively curved surfaces.  

Several previous studies \cite{Grason2007a, Grason2009, Grason2010a, Grason2012, Bruss2012a, Bruss2013, Hall2016} have investigated how each of these ``morphological responses" may effectively lower the cost of geometric frustration imposed by chiral order in bundles. Here we address how these distinct responses compete in the broader phase space of chiral filament assembly.  Specifically, we use a combination of continuum elasticity theory and discrete-filament bundle simulations to show that equilibrium morphology is selected by a combination of key thermodynamic (bundle surface energy), geometric (preferred chiral twist) and mechanical (stiffness of filaments and their interactions) properties of filament assemblies.  In Sec. ~\ref{sec: continuum}, we present a generic continuum model of defect-free twisted bundle assemblies, which allows us to compare the structure and thermodynamics of finite width bundle assemblies (possessing both isotropic and anisotropic cross sections) to the cost of unwinding chiral interactions to allow bulk order to grow without frustration.  We show that the possibility of ``escaping frustration" is governed by the ratio of stiffness of chiral filament interactions to the combined surface energy and mechanical costs of forming finite width twisted domains.  Above a critical chiral interaction stiffness, bundles exhibit both distinct isotropic (cylinder) and anisotropic (tape-like) morphologies, while below this critical stiffness, finite-diameter bundles are favored only within a narrow range of weakly-cohesive surface energies, above which equilibrium bundle growth causes assemblies to untwist to the parallel, unfrustrated state.   In Sec. ~\ref{sec: discrete}, we present discrete filament bundle simulations that compare the relative stability of bundles with anisotropic lateral dimensions to bundles which possess topological defects in their interior order.  We find that the optimal morphological response (boundary shape or bulk defects) is determined by the filament number $N$, the bundle twist and the ductility of inter filament bonds.  For small-$N$ and ductile inter-filament interactions, bundles pass directly from defect-free cylinders (i.e. isotropic boundary shape) to defective cylinders with increasing twist, while for large-$N$ and brittle interactions, the defective cylinder phase is preempted by an intermediate phase of defect-free anisotropic (tape-like) bundles.  We argue that the stability window of anisotropic tapes can be modeled via continuum theory estimates for the relative costs of straining inter-filament bonds (by geometric frustration) compared to the surface energy cost of introducing excess (anisotropic) boundary.  We conclude with a discussion of these results in the context of outstanding questions regarding the role of geometric frustration in shaping the assembly of chiral filament bundles.

\section{Escaping frustration to bulk assembly: continuum model of morphology selection}

\label{sec: continuum}

In this section, we present a continuum elastic theory for the formation of twisted bundles of chiral filaments that relates the free energy of assembly to bundle morphology, specifically the equilibrium twist and lateral dimensions of the cross section.  We begin by introducing the model and its application to the thermodynamics of assembly with isotropic versus anisotropic bundle cross sections, and then analyze the possibility of bundle assemblies to ``escape frustration" by relaxing inter-filament twist with increasing size. 

\subsection{Model}

\begin{figure}
\centerline{
\includegraphics[width=0.75 \linewidth]{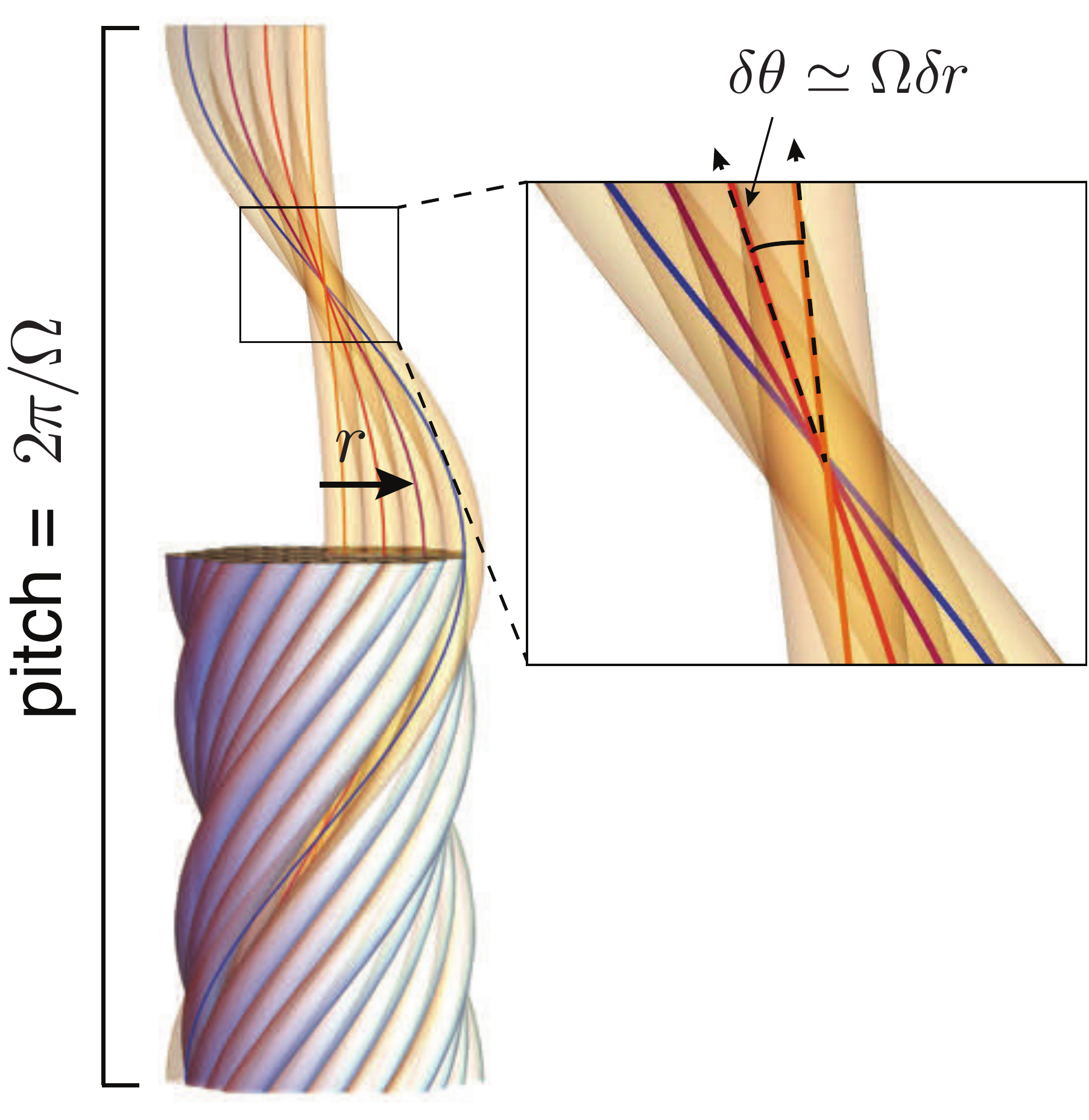}
}
\caption{\label{Figure 2}
The double-twist geometry is characterized by twist rate $\Omega$, related to the helical pitch of filaments and the inter-filament skew angle, $\delta \theta$. Here, $r$ describes the radial distance from the central axis.
}
\end{figure}

We describe the assembly of twisted bundles as finite domains of filaments with 2D columnar order and preference for chiral textures of filament backbones~\cite{Grason2012}. A columnar bundle is described by the shape of its 2D cross section normal to the mean-orientation of filaments ($\hat{z}$ axis), a local 2D displacement ${\bf u} (\xv)$ ($\perp$ to $\hat{z}$) of local filament positions away from a parallel and equally-spaced (e.g. hexagonal) reference configuration, and the unit vector orientation field of filament backbones ${\bf t} (\xv) =(\hat{z} + \partial_z {\bf u})/\sqrt{1+|\partial_z {\bf u}|^2}$.  Given the shape and geometry of the bundle (described by ${\bf u} (\xv)$ and  ${\bf t} (\xv)$) the following free energy describes bundle formation
\begin{equation}
\label{eq: Ftot}
F = F_{assem}+F_{elast} + F_{chiral}.
\end{equation}
Here,
\begin{equation}\label{fassem}
F_{assem} = - \mu \rho_0 A L + \Sigma P L ,
\end{equation}
represents the bulk and surface terms for a bundle of length $L$, cross sectional area and perimeter, $A$ and $P$, respectively, and areal filament density $\rho_0$.  The net per filament gain of inter-filament cohesion (minus entropic loss) to add filaments to a bundle is $-\mu$, while fewer cohesive interactions at the lateral surface of the bundle leads to a positive surface energy $\Sigma$~\footnote{We assume the length of bundles to be sufficiently long to neglect cost of ``surface slip" between neighbor filaments at the bundle ends.}.  The elastic energy has the form,
\begin{equation}
\label{eq: elastic}
F_{elast} = \frac{ B \rho_0}{2} \int dV ~ \kappa^2(\xv) + \frac{1}{2} \int dV ~\sigma_{ij} u_{ij} ,
\end{equation}
where the first term represents the cost for intra-filament, bending deformation, with $B$ and $\kappa (\xv) = |\partial_z \tv|$ the filament stiffness and curvature, respectively.  The second term describes the continuum elastic cost of deforming lateral inter-filament order (spacing and symmetry) where  $\sigma_{ij}$  and $u_{ij}$ are the respective in-plane stress and strain tensors (with components in the plane of lattice order, $\perp$ to $\hat{z}$).  We assume a isotropic, linear elasticity appropriate for hexagonal inter-filament order $u_{ij} = Y^{-1} \big[(1+\nu) \sigma_{ij} - \nu \delta_{ij} \sigma_{kk}\big]$ where $Y$ and $\nu$ are the respective Young's modulus and Poisson's ratio of the 2D filament array.  The geometric coupling between positional and orientational degrees of freedom is built into the form of the non-linear columnar strain~\cite{Selinger1991},
\begin{equation}
u_{ij} = \frac{1}{2}\big( \partial_i u_j+ \partial_j u_i - t_i t_j\big).
\end{equation}
The non-linear contribution from tilt is necessary to preserve rotational invariance of the theory, and more intuitively, represents the fact the distance of closest approach between filaments will be {\it reduced} when filaments tilt relative to $\hat{z}$ (i.e. $t_i \neq 0$) without changing their positions in that plane (i.e. $\partial_i u_j =0$).   The final contribution to $F$ derives from orientation dependent interactions between chiral, rod-like elements, and hence, has the form of the twist-elastic contribution to the Frank energy of nematic phases \cite{deGennes1995physics}, 
\begin{equation}
F_{chiral} = \frac{K}{2} \int dV ~\big[ \tv \cdot (\grad \times \tv) + q_0 \big]^2,
\end{equation}
where $K$ is the twist elastic constant (often written as $K_{22}$) and $q_0\neq 0$ parameterizes the drive of chiral filaments to adopt a twisted texture with $ \tv \cdot (\grad \times \tv) = -q_0$~\footnote{Note the bend-elastic term is already included in eq. (\ref{eq: elastic}).  In columnar materials, in-place positional order strongly suppresses splay deformations, i.e. $\grad \cdot \tv \neq 0$. For ``double-twisted" textures considered here, splay is strictly zero, and hence we omit any explicit splay elastic cost for simplicity.}.

In simple chiral nematics (which lack positional order) $q_0\neq 0$ leads to a uniaxial cholesteric ground-state texture, with nematic director rotating along a direction orthogonal to the nematic axis with a pitch $2 \pi/q_0$.   Due to the 2D positional order, uniaxial cholesteric textures require constant longitudinal gradients of in-plane shears~\cite{Kamien1996}, and hence, are suppressed as $L \to \infty$.  Hence, for finite chiral columnar domains such as bundles, chirality favors an alternative ``double twist" texture~\cite{Wright1989a}, in which filament orientations wind helically around a central axis, and filament positions are simply rigidly-rotated (shear-free to linear order) from one transverse section to the next along the bundle.  Parameterizing this texture by the {\it twist} $\Omega$ of filaments around the axis, the backbone texture is 
\begin{equation}
\label{eq: double}
\tv (\xv) \simeq \hat{z} + \Omega r \hat{\phi} ,
\end{equation}
where $(r,\phi)$ are polar coordinates with respect to the twist axis.  In this geometry, 1) each filament adopts a helical shape and possesses a single pitch $P = 2 \pi/\Omega$; and 2) neighbor filaments spaced $\delta r$ along the radial direction adopt a nearly constant inter-filament skew angle $\delta \theta \simeq \Omega \delta r$ (see Fig. 2).  Because $\tv \cdot (\grad \times \tv) \simeq 2 \Omega \propto \delta \theta$ for this texture, the preference for double-twist in the bundle can therefore be linked to microscopic interactions between chiral filaments which favor a fixed inter-filament skew between neighbors $\delta \theta_0 \propto q_0$ and generate an inter-backbone torque $\propto K(\delta \theta - \delta \theta_0)$ for deviations from this preferred local packing for which $F_{chiral}/V \simeq 2 K(\Omega-\Omega_0)^2$, where $\Omega_0 \equiv - q_0/2$ is the rate of bundle twist preferred by chiral interactions. \cite{Grason2009,Grason2007a}

Given this double twist texture, these elastic costs for intra-filament and inter-filament deformations may be derived for a given 2D cross-section.  For example, since filament curvature grows linearly with distance from the bundle axis $\kappa (\xv) = \Omega^2 r$, the bending energy (per unit volume) takes the simple form $F_{bend}/V = B \rho_0 \Omega^4 \langle r^2 \rangle/2$, where $ \langle r^2 \rangle$ is the moment of interia of the cross section.  Calculation of the inter-filament elastic energy requires solution of the mechanical equilibrium equations, which are the Euler-Lagrange equations for $\uv (\xv)$, $\partial_i \sigma_{ij} = 0$ combined with a compatibility condition that relates in-plane gradients of inter-filament stresses to in-plane gradients of filament tilt $\tv_\perp$ (in-plane components of $\tv$)
\begin{equation}
Y^{-1} \grad_\perp^2 \sigma_{ii} \simeq - K_{eff} ,
\end{equation}
where $K_{eff} \simeq \grad_\perp \times \big[\tv_\perp (\grad_\perp \times \tv_\perp) - (  \tv_\perp \times \grad_\perp ) \tv_\perp \big]/2$ and plays a role equivalent to the Gaussian curvature in 2D elastic membranes, connecting gradients of out-of-plane tilt to gradients of in-plane stress.  For the double-twist texture in eq. (\ref{eq: double}) $K_{eff} = +3 \Omega^2$, where the small-tilt approximation assumes sufficiently large helical pitch such that $(\Omega r) \ll 1$, the frustration of inter-filament spacing is equivalent to crystalline order on a sphere of radius $|\Omega^{-1}|/\sqrt{3}$~\cite{Grason2015}. 

In this model, consider two limiting classes of bundle cross-sectional shape:  {\it cylindrical} and {\it tape-like} twisted bundles.  Cylindrical bundles possess an isotropic (circular) cross-section, which would be optimal from the point of view of surface energy.  For a cylindrical bundle of radius $R$ the inter-filament elastic pressure has the form, $\sigma_{ii} = 3 Y \Omega^2(R^2-2r^2)/8$, illustrating the tendency of compression to increase with distance from the bundle center, analogous to the geometric compression of a (flat) 2D membrane confined to the spherical surface.  From this stress distribution and the second-moment of the cross section, $\langle r^2 \rangle =R^2/2$, we have 
\begin{equation}\label{felascyl}
F_{elast, cyl}(R)/V = \frac{\rho_0 B \Omega^4 R^2}{4} + \frac{ 3 Y (\Omega R)^4}{128}.
\end{equation}
In the second morphology, the helical tape, we assume the cross-section has a rectangular shape with one lateral dimension, the width $w$, much larger than its transverse thickness $t$.  We approximate the stress distribution by taking the $w\gg t$ limit, where introducing Cartesian coordinates $(x,y)$ along the respective width and thickness axes, stresses will be independent of $x$ (away from distant ends at $x=\pm w/2$).  In this case, mechanical equilibrium and vanishing normal stress along the wide edges require $\sigma_{yy} \simeq \sigma_{xy}\simeq 0$.  Combining compatibility with the condition of zero average stresses along $\hat{x}$ direction, we find $\sigma_{xx} \simeq Y \Omega^2(t^2-12y^2)/16$.  Notably, this solution shows that compression increases away from the line $y=0$, leading to stress growth only along the {\it thin} dimension of the cross-section (as opposed to $w$).  This illustrates the effect of introducing additional {\it free surface} to the frustrated assembly.  Compatibility requires a nonzero stress drop across that bundle cross section, and the introduction of free surface allows that stress drop to take place along the narrow dimension of the domain through unconstrained displacement of the boundary, thereby lowering the elastic cost of frustration (at the expense of breaking cohesive bonds and raising the surface energy).  Integrating the elastic stress through the section and using $\langle r^2 \rangle = (w^2+t^2)/12$ we have have
\begin{equation}\label{felastape}
F_{elast, tape}(w\gg t)/V = \frac{\rho_0 B \Omega^4 (w^2+t^2)}{24} + \frac{  Y (\Omega t)^4}{160} 
\end{equation}
This shows the key result that elastic costs (per unit volume) of {\it extrinsic} (i.e. bending) geometry of anisotropic bundles are most sensitive to the {\it wide} dimension (growing as $\sim w^2$) as compared to costs of {\it intrinsic} (i.e. metric) geometry which are controlled by the narrow dimension (growing as $\sim t^4$).  

To analyze the free energy density of competing morphologies it is useful to introduce two material dependent length scales:
\begin{equation}
\lambda_S \equiv \Sigma/Y ; \ \lambda_B \equiv \sqrt{\frac{ \rho_0 B}{Y} } ,
\end{equation}
where the first length scale, the {\it elasto-cohesive length}, effectively characterizes the ratio of depth and elastic stiffness of inter filament bonds, and the second, called the {\it bend penetration depth} in the context of columnar order~\cite{deGennes1995physics}, characterizes the relative cost of intra-filament bending in comparison to inter-filament bond stretching.  Further by introducing the dimensionless length scales, 
\begin{equation} \label{dimensionlessdefs} 
\rho \equiv \frac{R}{\lambda_B}; \upsilon \equiv \frac{w}{\lambda_B}; \tau \equiv \frac{t}{\lambda_B}; \omega \equiv \frac{\Omega}{\Omega_0};
\end{equation}
and energy scales,
\begin{equation}
s \equiv \frac{\Sigma}{Y \Omega_0^4 \lambda_B^5}; k \equiv \frac{K}{Y \Omega_0^2 \lambda_B^4}; f\equiv \frac{F}{V Y (\Omega_0 \lambda_B) ^4} 
\end{equation}
we arrive at dimensionless free energies for cylindrical 
\begin{equation}
f_{cyl} = \frac{2 s}{\rho}+2 k(\omega-1)^2 +\frac{\omega^4 \rho^2}{4} + \frac{3 (\omega \rho)^4}{128}
\end{equation}
and tape-like bundles
\begin{equation}
f_{tape}= 2 s \big(\tau^{-1}+\upsilon^{-1}\big)+2 k(\omega-1)^2 +\frac{\omega^4 \upsilon^2}{24} + \frac{(\omega \tau)^4}{160}
\end{equation}
where we neglect the $\tau^2$ contribution to bending in the $w \gg t$ limit and we have dropped the constant contribution from the bulk chemical potential.

\begin{figure*}
\centerline{
\includegraphics[width=1.0 \linewidth]{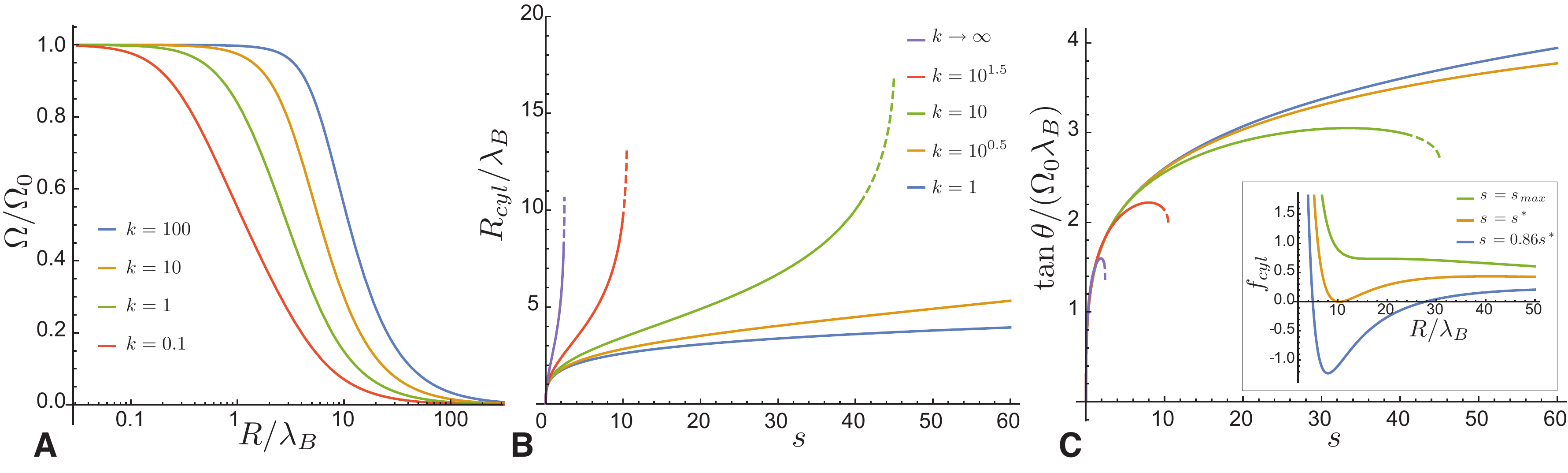}
}
\caption{\label{Figure 3}
In (A), plotting equilibrium twist of cylindrical bundles $\Omega$ (relative to value preferred by chirality, $\Omega_0$) as a function of outer bundle radius, for a range of reduced chiral stiffness, $k= K/(Y \Omega_0^2 \lambda_B^4)$, shows untwisting radius (defined by $\Omega= \Omega_0/2$) increases with $k$.  In (B), we show the selection of equilibrium cylinder size, $R_{cyl}$, as a function of reduced surface energy $s= \Sigma/(Y \Omega_0^4 \lambda_B^5)$, for a range of reduced chiral stiffness $k$. In (C), the outer filament twist angle $\tan \theta = \Omega R_{cyl}$ of equilibrium cylinders, shows a generic non-monotonic dependence on surface energy.  The inset in (C), shows the radius dependence of twisted cylinders near to the transition from finite-diameter bundles and bulk, untwisted assembly at $s = s*$ for $k=10$.  In (B) and (C), dashed curves show metastable branches (for $s* <s<s_{max}$) where finite cylinders are separated from (equilibrium) bulk assembly by a finite free energy barrier.
}
\end{figure*}

\subsection{Cylindrical Bundles: Finite Diameter vs. Frustration Escape}

First, we analyze the equilibrium conditions for bundles assumed to form cylindrical morphologies.  Specifically, we analyze the role played by the strength of chiral inter-filament interactions, as parameterized by $k$, to control the ability to unwind the helical twist of the bundle (i.e. $\omega <1$) and allow the bundle to reach larger equilibrium sizes than possible for fixed twist, $\omega=1$.  Here, we consider the canonical ensemble and assume conditions (e.g. saturated filament solutions) where all but a negligible fraction of filaments are condensed into bundles.  In this case, assuming bundles adopt a mean radius $\rho$ (with negligible size fluctuations), the preferred size and twist of bundles follows from minimization of $f_{cyl}$ with respect to $\rho$ and $\omega$ yielding the equations of state, respectively
\begin{equation}
\label{eq: req}
s = \frac{\omega^4 \rho^3}{4} + \frac{3 \omega^4 \rho^5}{64},
\end{equation}
and
\begin{equation}
k = \omega^3\bigg( \frac{ \rho^2 + 3 \rho^4/32}{4(1-\omega)} \bigg).
\end{equation}
Solving the latter cubic equation for equilibrium twist we find,
\begin{equation}
\label{eq: omcyl}
\omega_{cyl} = \frac{3}{2 \Gamma} \Big[x^{1/3}(\Gamma) - x^{-1/3}(\Gamma) \Big] ,
\end{equation}
where $x(\Gamma) = \Gamma + \sqrt{1+\Gamma^2}$ and 
\begin{equation}
\Gamma^2(k,\rho) = \frac{27}{16 k} \rho^2(1+3 \rho^2/32).
\end{equation}
Fig. 3A shows the radius dependence of equilibrium twist.  Small bundles adopt preferred twist, $\omega_{cyl} \simeq 1$, while large bundles untwist as $\omega_{cyl} \sim (k/\rho^4)^{1/3}$ as $\rho \to \infty$ due to the size dependent elastic costs of twist frustration.  The crossover between ``twist-locked" and ``mechanically-untwisted" bundles can be associated by the ``half-twist size" $\rho_{1/2}$ defined by $\omega_{cyl} (r_{1/2})=1/2$ in eq. (\ref{eq: omcyl}).  For weak chiral interactions ($k\ll 1$) $\rho_{1/2} \sim  k^{1/2}$ indicating that bundles prefer to untwist upon reaching only a small radius, while for stiff chiral interactions ($k \gg 1$) $\rho_{1/2} \sim k^{1/4}$ indicating that in the limit of clamped chirality ($k \to \infty$) bundles retain preferred twist at all sizes.  

Inserting the size-dependent twist, eq. (\ref{eq: omcyl}), into eq. (\ref{eq: req}) yields the relation between surface energy and equilibrium bundle size $\rho_{cyl}$.  In the limit of $s \ll 1$, bundles are driven by relatively weak cohesive interactions and retain preferred twist ($\omega_{cyl} \simeq 1$), reaching only a narrow size $\rho_{cyl} \sim s^{1/3}\ll 1$ limited by intra-filament bending.  The growth of bundles with increasing $s$ is highly dependent on the stiffness of chiral interactions, $k$ (see Fig. \ref{Figure 3}B).  In the  $k \to \infty$ limit of ``twist-locked" interactions, bundles crossover from $s \ll 1$ to size limited by geometrically-induced inter-filament stress  at large surface energy, $\rho_{cyl} \sim s^{1/5} \gg 1$.  For finite $k$, the reduction of equilibrium twist (shown in Fig. \ref{Figure 3}A) leads to larger bundle sizes for a given $s$, relative to the ``twist-locked" limit, due to the untwisting of bundles with increasing size.  This leads to a maximal $s$ beyond which surface energy drives the bundle to completely untwist and reach infinite size.  Thus, assembly can escape the costs of geometric frustration provided that the cohesive drive (as quantified by $s$) is sufficiently strong to untwist assembly to the parallel state.  This follows from the fact that beyond $r_{1/2}$ bundle twist falls rapidly, and consequently the size-dependent mechanical costs fall (proportional to $\omega_{cyl}^4$) allowing bundles to grow to much larger sizes, peeling away from the $k \to \infty$ behavior and ultimately exhibiting a singular divergence at a critical surface value $s_{max}$.  The $k$-dependence of this maximal $s_{max}$  can be estimated by using half-twist condition ($\rho_{1/2}$ and $\omega_{cyl}=1/2$) in the equation of state, eq. (\ref{eq: req}), yielding $s_{max} \sim k^{3/2}$ for $k\ll1$ and $s_{max}\sim k^{5/4}$ for $k \gg 1$.  Thus, the stable range of self-limiting chiral filament bundle assembly vanishes as $k \to 0$, and conversely, extends at all $s$ for infinitely stiff chiral interactions.

Before considering the effect of ``frustration escape" on the competition between cylindrical and tape-like bundles, we point out two features of the untwisting transition.  Fig. \ref{Figure 3}C shows the non-monotonic evolution of the twist angle $\theta = \arctan (\Omega R)$ as a function of $s$ for a range of chiral stiffness, $k$.  For weakly cohesive assembly, $\theta$ generically increases with $s$ due to the increasing radius and roughly constant pitch, while for sufficiently large $s$, bundles untwist substantially (say for $s$ near the half-twist condition) leading to a maximum $\theta$ beyond which twist angle begins to decrease before a singular drop at $s_{max}$.  A second feature is illustrated in the inset of Fig. \ref{Figure 3}C, showing plots of the free energy vs. bundle radius near to $s_{max}$, where $f_{cyl}$ is plotted relative to the bulk, untwisted ($\omega \to 0, \rho \to \infty$) state.  As surface energy approaches this value from below, the depth of the finite-$\rho$ minimum increases until it reaches a critical value $s^*_{cyl}<s_{max}$ where finite cylinder assembly is in equilibrium with bulk, untwisted assembly.  For $s^*_{cyl}<s \leq s_{max}$ finite diameter bundles remain metastable in the regime of equilibrium bulk order (depicted as dashed curves in Fig. \ref{Figure 3}), while for $s\geq s_{max}$ the barrier associated with untwisting to the parallel, unfrustrated state is removed.  While bundle metastability is a generic feature for finite $k$ assembly, the range of metastable $s$ is relatively narrow, because $s_{cyl}^*$ exhibits the same scaling dependence on stiffness of chiral interactions as $s_{max}$.

\subsection{Twisted Cylinders vs. Helical Tapes vs. Bulk, Untwisted Assembly}

We now revisit a previous prediction for morphology transition between twisted cylindrical bundles and twisted tapes, which assumed fixed twist (i.e. fixed $\Omega$) as a opposed to fixed chiral interactions (i.e variable $\Omega$ for fixed $K$ and $\Omega_0$).  Previously, we showed that for fixed $\Omega$, as $\Sigma$ increases, the rapidly growing cost of inter-filament frustration drives a morphology transition to highly anisotropic tapes.  In this case, frustration by chirality is fixed, so that while tapes reach lateral dimensions (widths) that are larger than would be possible for cylindrical morphologies, they remain finite for all $\Sigma$ due to the non-vanishing cost of finite twist.  Hence, in the present model (fixed chiral interactions) chirality acts as a ``soft frustration" mechanism, which can be overcome for sufficiently strong cohesive assembly drives by untwisting from the preferred chiral twist to parallel assembly.   

Based on the continuum model, we consider the competition between three phases:  cylindrical bundles (finite $R$ and $\Omega$), tape-like bundles (finite $t, w$ and $\Omega$), and untwisted bulk assembly ($R \to \infty$ and $\Omega \to 0$).  Along with the equations of state for cylindrical bundles, the free energy of tape-like bundles is determined via the solution for the equations of state of lateral bundle dimensions and twist (minimization of $f_{tape}$ with respect to $\tau$, $\upsilon$ and $\omega$),
\begin{equation}
s = \frac{\omega^4 \upsilon^3}{24}=\frac{\omega^4 \tau^5}{80} ,
\end{equation}
and 
\begin{equation}
k=\omega^3\Big(\frac{\upsilon^2+3\tau^4/20}{ 24(1-\omega)}\Big).
\end{equation}
The first two conditions are easily solved for a given twist $\upsilon=(24 s / \omega^4)^{1/3}$ and $\tau=(80 s / \omega^4)^{1/5}$. When $s \gg \omega^{4}$, due to the weaker growth of the bending energy ($\sim \upsilon^2$) compared to the cost of inter-filament frustration ($\sim \tau^4$), large cohesive energies cause tapes to grow increasingly more anisotropic, leading to $w/t \gg 1$.  

Solving the additional condition for twist equilibrium, we find a qualitatively similar trend as described above for isotropic cylinders:  tape dimensions increase with approximately constant $\omega_{tape} \simeq 1$ for small $s$, and at a critical size range equilibrium tapes rapidly unwind leading to a singular increase in tape size at critical $s$, the value of which increases with $k$.  Indeed, because the thickness of tapes scales in the same way as cylinder radius for large $s$, the escape to bulk untwisted assembly ($\omega_{tape} \to 0$) occurs at comparable, if somewhat larger, values of surface energy for tapes relative to cylindrical bundles.

\begin{figure}
\centerline{
\includegraphics[width=0.95 \linewidth]{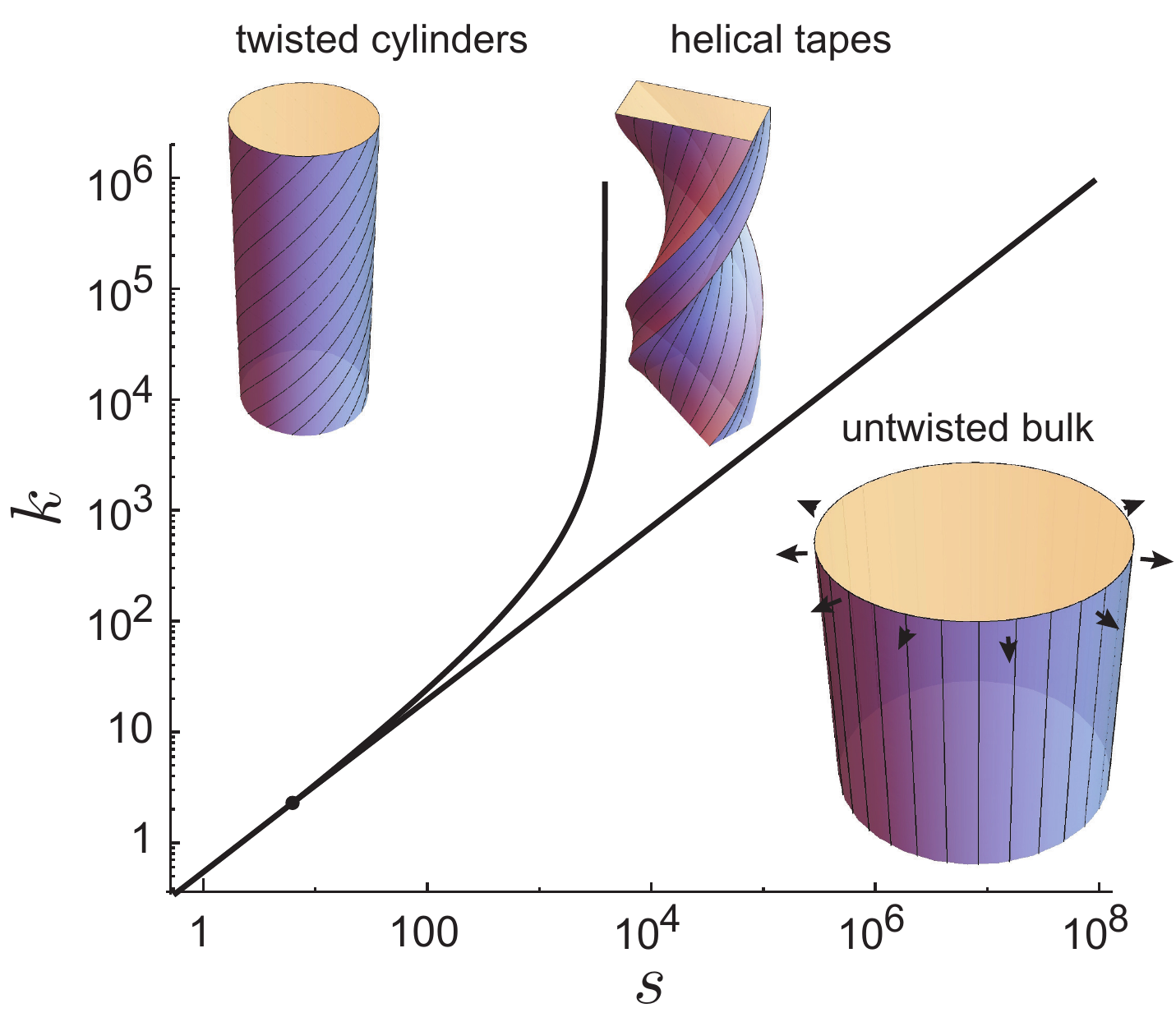}
}
\caption{\label{Figure 4}
Equilibrium morphology phase diagram for condensed chiral filaments (canonical ensemble) according to stiffness of chiral interaction $k= K/(Y \Omega_0^2 \lambda_B^4)$ versus surface energy $s= \Sigma/(Y \Omega_0^4 \lambda_B^5)$.   Filled circle indicates a triple point at $s_{tri}=6.65$ and $k_{trip}=2.4$, where finite-cylinders, tapes and bulk, untwisted assemblies coexist.
}
\end{figure}

The boundary between equilibrium cylinders and tapes is found by solving the condition $f_{cyl}=f_{tape}$ for equilibrium structures, while the transition to bulk (untwisted) assembly is determined by comparing free energies of bundle morphologies to $f_{bulk} = 2 k$.  For the $k \to \infty$ limit of stiff chirality we find that reduced inter-filament frustration, at the expense of excess surface energy, in tapes leads to a transition from cylinders to tapes at $s_{c/t} (k \to \infty) \simeq 3,860$, as previously reported in ref. \cite{Hall2016}.  For finite $k$, soft chiral interactions allow bundles, both tapes and cylinders, to untwist to bulk parallel assembly above a critical surface energy, as estimated by $s_{max}$ above.  Whether this untwisting transition occurs from a cylinder phase or tape phase of bundles can crudely be determined by comparing the $k$ dependence of $s_{max}$ to $s_{c/t}$.  This crude estimate suggest a critical range of $k\approx 100$, above which stiff chiral interactions allow bundles to grow to sufficient size that inter-filament frustration drives a morphology transition to equilibrium tapes at large $s$.  For ``soft" chiral interactions ($k \lesssim 100$), frustration escape proceeds at sufficiently low $s$ to prevent a broad range of stable equilibrium tapes before the bulk phase is reached at high $s$

This crude picture is consistent with the 3-phase assembly phase diagram computed from exact solution of the equations of state above, as shown in Fig. \ref{Figure 4}.  We note, however, that even in the regime of ``soft" chiral interactions ($k \lesssim 100$) a narrow window of equilibrium persists between cylindrical bundles and bulk assembly due to the fact that bundle sizes grow dramatically with $s$ near to the untwisting transition, generically accompanied by a sufficient rise in inter-filament frustration costs to drive a transition to tapes, albeit tapes which are also very close to the point of frustration escape via untwisting.  A narrowing window of tapes persists down to a {\it triple point} at $s_{tri}=6.65$ and $k_{tri} = 2.4$, below which very soft chiral interactions permit weak surface energy to drive finite cylinders directly to bulk, untwisted filament arrays.

\section{Reshaping the boundary vs. reorganizing the inter-filament order: discrete-filament model of morphology}

\label{sec: discrete}

In this section, we investigate how frustrated chiral bundle assemblies select between two fundamentally distinct morphologies: defect-free bundles with anisotropic shapes (tapes) and bundles possessing topological defects in the interior, but with otherwise isotropic cross sections.  Previous studies have investigated these distinct morphologies as independent structural responses to frustration in chiral bundles.

Disclinations are point-like locations in a 2D quasi-hexagonal packing where 6-fold inter-filament bond order breaks down.  As shown schematically in Fig. \ref{Figure 1}E, 5-fold disclinations allow partial relaxation of the frustration-induced compression along the azimuthal direction in twisted bundles \cite{Grason2010a, Grason2015}, akin to how they relax hoop compression in spherically-curved crystals \cite{Azadi2016}.  Defect configurations are characterized by their net topological charge $Q$, which is the sum of total deficit in coordination (e.g., assigning value +1/-1 to 5-fold/7-fold disclinations).  Previous theory \cite{Grason2010a} and simulation studies \cite{Bruss2012a, Bruss2013} of twisted bundles with quasi-cylindrical boundary shapes show that the stability of excess disclinations in the ground state packing is controlled by the product $\Omega R\equiv \tan \theta $, which corresponds to the area integral of Gaussian curvature of packings mapped onto the dual surface \cite{Grason2015}.  For bundles $\theta \leq \theta_c = \arctan(\sqrt{2/9}) \simeq 0.44$, cylindrical bundles favor $Q=0$ packings (largely defect free), while for $\theta > \theta_c$ ground states possess {\it excess 5-fold} disclinations, $Q>0$ with a number that increases from +1 to a maximum of $+6$ for highly twisted bundles ($\theta \to \pi/2$).  

In a parallel study, Hall and coworkers \cite{Hall2016} investigated the possibility for bundles to relax frustration without defects, {\it through a reorganization of boundary shape}.  As described above, surface energy drives bundles to form larger diameters, which in turn increases the costs of frustration-induced elastic energy.  Because the cost of metric frustration grows rapidly with the area of isotropic bundle cross sections $A$, growing as $\sim A^2$, relatively stiff inter-filament bonds ultimately lead to an elastic instability of boundary shape that drives cross sections to adopt anisotropic shapes (i.e. tape-like morphologies).  This instability, which parallels the behavior of defect free crystal growth on spherical surfaces \cite{Meng2014, Schneider2005, Kohler2016}, lowers the elastic cost of inter-filament frustration by limiting the {\it narrow} dimension of the structure, at the expansion of increased surface energy and intra-filament bending cost.  In this previous study, which considered the (fixed twist) canonical ensemble, we showed that this bundle to tape transition occurred at the critical value of surface energy, $s_c \simeq 3,860$ (simulation results showed a lower threshold, $s_c \approx 20-60$).

Here, we employ a discrete filament bundle simulation model to compare the relative stability of these competing morphological responses.  For simplicity, we focus this study on the {\it fixed twist} and {\it fixed filament number} (microcanonical) ensembles.  In contrast to previous studies, we compare numerically simulated bundle cross sections that alternately optimize {\it boundary shape} and {\it inter-filament order}, and investigate the conditions that select between boundary shape vs. defect-mediated morphology response.

\begin{figure*}
\centerline{
\includegraphics[width=0.95 \linewidth]{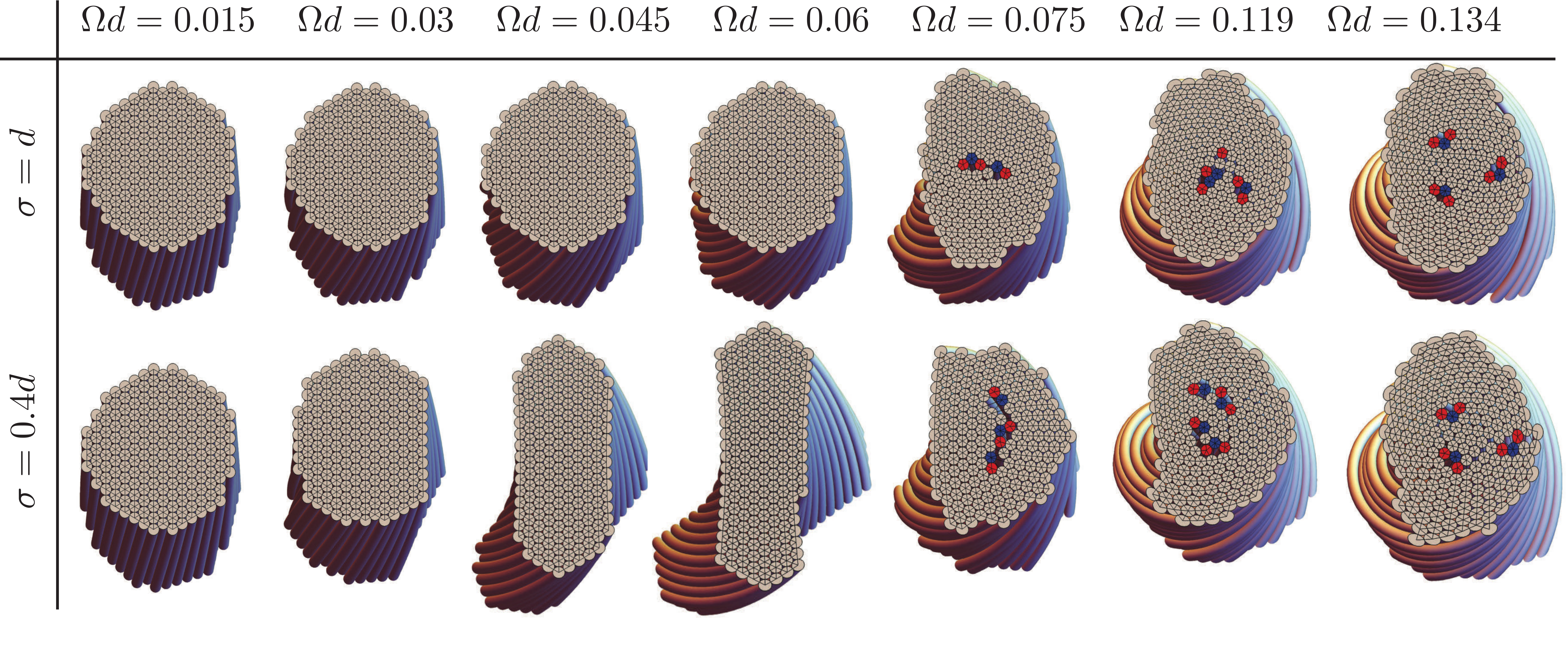}
}
\caption{
Simulated ground states of discrete-filament model of chiral bundles at fixed $N = 181$, for increasing values of reduced twist $\Omega d$, where $d$ is equilibrium center-to-center filament spacing, or diameter.  Top row shows a relatively ductile case where cohesive interaction range $\sigma$ is equal to $d$, showing a transition from (approximately) cylindrical defect-free bundles to defective, cylindrical packings possessing excess 5-fold disclinations, i.e. $Q = +1, +2$ and $+3$.  Red (blue) filaments label sites of interior 5-fold (7-fold) disclinations.  The bottom row shows a relatively ductile case, where low-twist, defect-free cylinders transition to intermediate-twist anistropic ``tape-like" bundles, followed by a re-entrent cylindrical morphology (with excess defects) at high twist.  }
\label{morphs181}
\end{figure*}

\subsection{Discrete filament bundle model and simulation methods}

We consider a discrete filament model of chiral filament bundles.  Bundles are described by a fixed number of filaments $N$ and fixed (post-equilibrium) value of twist $\Omega$.  The energy of bundle assembly is modeled by the energy of a cross-sectional slice of height $\Delta z$, whose energy $\Delta E$ is described by
\begin{equation}
\label{deltaE}
\Delta E = \sum_{i=1}^{N} \bigg(\frac{B}{2} \kappa_i^2 + \sum_{ \rho_j \leq \rho_i} \frac{u(| \bD_{ij} |) }{\sqrt{1+ \kappa_j \bD_{ij} \cdot {\bf N}_j }} \bigg) \Delta \ell_i.
\end{equation}
The first term describes bending energy of filaments with $\kappa_i = \Omega^2 \rho_i/[1+( \Omega \rho_i )^2 ]$ with $\rho_i$ the radial distance of the $i$th filament from the twist axis.  The second term describes the cohesive interaction of filament $i$ with filament $j$ (by definition at an inner radial distance, $\rho_j \leq \rho_i$), which can be captured by the distance of closest approach $\bD_{ij}$ between center lines of the filaments.  Due to the 3D geometry of filament backbones, the closest contact between curves of $i$ and $j$ is generically out of the plane of the cross section, depending not only on distance between points in the cross section, but also on their distance from the central axis and the bundle twist $\Omega$ (see eqs. \ref{eqDOCA} and \ref{eqDOCA2} in appendix \ref{subs Discrete filament model}).  Here, we choose the same class of cohesive potentials as in ref. \cite{Hall2016}
\begin{equation}
\label{uDelta}
u(\Delta) = \frac{\epsilon}{6}\Big[\frac{ 5 \sigma^{11}}{(\Delta + \sigma - d)^{11} }- \frac{ 11 \sigma^{5} }{(\Delta + \sigma - d)^5 } \Big]
\end{equation}
where $\epsilon$ is the cohesive energy per unit length of filaments separated at an equilibrium distance $d$ (e.g., their diameter).  We choose a ``5-11" potential since the form follows directly from the integration of a pair-wise Lennard-Jones potential between a point on $i$ and a (locally straight) section of $j$, $\Delta_{ij}$ away, where the factor of $( 1+ \kappa_j  \bD_{ij} \cdot {\bf N}_j )^{-1/2}$ corrects for the curvature of $j$ (with ${\bf N}_j$, its normal). Here, $\sigma$ sets the ductility of pairwise bonds: $\sigma \ll d$ results in brittle inter-filament bonds and $\sigma \gg d$ results in ductile bonds.  In eq. (\ref{deltaE}) $\Delta \ell_i = \Delta z \sqrt{ 1+ ( \Omega \rho_i )^2}$ is the length element of $i$ in the cross section.  From the energy per slice, we define the (intensive) energy per filament length,
\begin{equation}
{\cal E} \equiv  \Delta E/   \Delta L ,
\end{equation}
where $ \Delta L = \sum_{i=1}^N \Delta \ell_i$ is the total filament length per cross-sectional slice. We focus simulations on the case of negligible bending energy content, and choose only a minimal non-zero value of $B= 10 \epsilon d^2$, so as to provide a slight inward force on filaments towards the bundle center.

To determine equilibrium morphologies for a given $N$, $\Omega$, bending stiffness and inter-filament potential, we minimize ${\cal E}$ with respect to in-plane filament coordinates $\xv_i$.   As the energy landscape of this optimization problem is rough due to the underlying frustration, resulting equilibria are highly dependent of both initial configurations and sampling protocol. Rather than seek a rigorous ground state (say via advanced sampling), here we aim to identify candidate ground state structures which are assumed to closely parallel the true ground state in terms of energy and structure.  As the structural competitors of ground states fall into at least two morphologically very distinct states (presumably separated by multiple large energy barriers) for each set of simulated bundle parameters we use a combination of the following methods.

{\it Method A} - This method considers filament positions which are initially defect-free and templated from an hexagonal lattice of spacing $d$, but fall within a cross-sectional ``stencil" of variable anisotropic structures.  The stencil is described by a stadium curve with lateral inner dimensions, $X_1$ and $X_2$, such that narrower ends of the stencil are capped by hemicircular arcs of radius $|X_1-X_2|/2$.  The stencil dimensions and boundary filaments are adjusted such that the total number of enclosed filaments is $N$ (see Appendix \ref{subs Simulation Details}), producing a distinct initial configuration that varies from nearly isotropic, cylinders ($X_1\approx X_2$) to highly anisotropic, tape-like bundles ($X_1 \gg X_2$).  From each of these distinct initial configurations, filament positions are first relaxed to avoid overlaps with $\Delta_{ij} < d$ (see Appendix \ref{subs Simulation Details}), and following this, relaxed according to conjugate gradient minimization of ${\cal E}$, and the lowest energy (per length) is selected as the defect-free ground state candidate.

\begin{figure*}
\centerline{
\includegraphics[width=0.9 \linewidth]{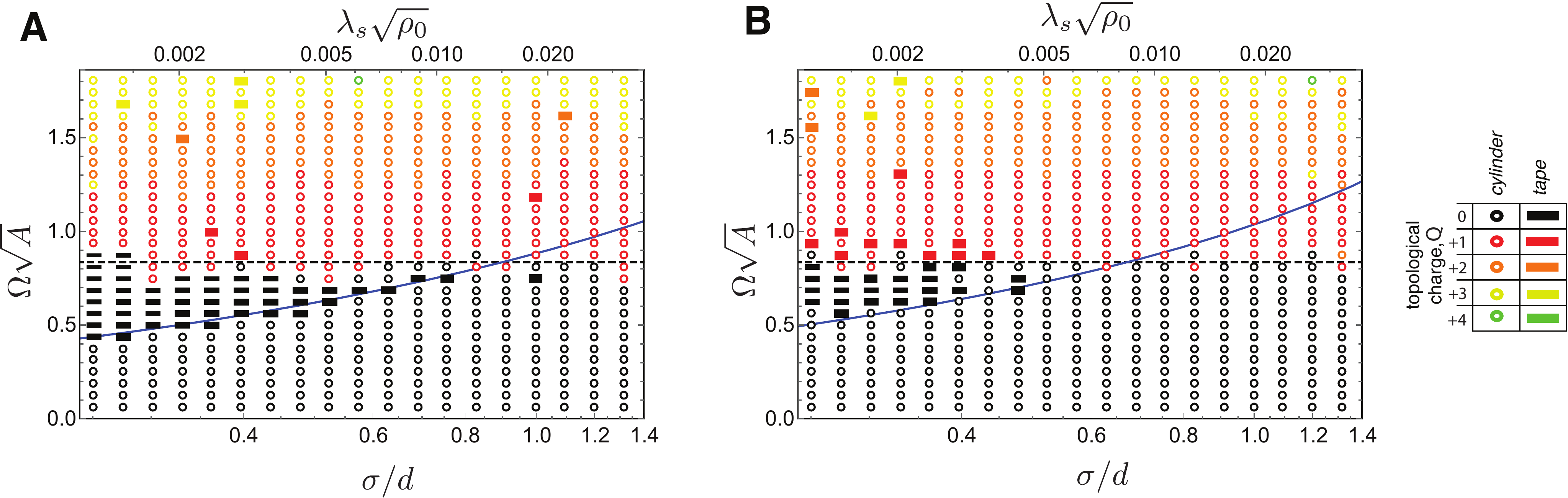}
}
\caption{
Morphology phase diagrams from discrete filament simulations for $N = 181$ (A) and $N=59$ (B), where a threshold value of aspect ration $\alpha = 1.5$ is used to distinguish cylinders from tapes and apparent aspect ratios of bars highlight optimal energy aspect ratios.  Dashed line is drawn from predicted threshold for critical twist to stabilize 5-folds in cylindrical bundles and solid blue line is determined by continuum theory prediction for fixed $N$ (and $\lambda_B=0$) transition between defect-free cylinders and tapes (as discussed in text). 
}
\label{phases}
\end{figure*}

{\it Method B} - This method makes no assumption of initial triangular order and instead allows bundles to form with uncontrolled defect types, numbers and locations in the bundle cross sections.  Bundle initial positions are generated by random placement of filament positions with a distance of $0.5 d \sqrt{N}$ from the bundle axis, followed by a post-initialization relaxation to remove overlaps (see Appendix \ref{subs Simulation Details}).  From random initial configurations, filament positions are  relaxed according to conjugate gradient minimization of ${\cal E}$.  For each set of bundle parameters, at least 3000  random initial configurations are generated, and the lowest final energy is selected as the ground state candidate.

Following application of both relaxation methods, ground state competitors are characterized according to lateral aspect ratio, $\alpha$, defined by the dimensions of the rectangle with minimal area that encloses the cross section, including filament volumes.  Additionally, the inter-filament packing is triangulated, making use of the conformal projection from the dual curved surface to the 2D plane \cite{Bruss2013}, {\it bulk} disclination defects are identified as positions of non 6-fold coordination not lying along the outer edge of the bundle, and the net topological charge $Q$ is computed as the sum of total deficit in coordination (relative to 6-fold) from that set of defects.

\subsection{Simulation Results and Discussion}

Results of minimal energy morphologies for $N=181$ filaments are shown in Fig.~\ref{morphs181} for two different interaction ranges  and a sequence of increasing twist, between $\Omega d = 0.015 -0.134$.  For the relatively ductile potential $\sigma=d$, we find that bundle cross sections remain roughly circular, but adopt excess 5-fold disclinations above a critical twist ($Q \geq +1$ for $\Omega d \geq 0.075$), and increasing topological charge with further twist, consistent with the results for the same potential in ref. \cite{Bruss2012a}.  In contrast, for the shorter-range, or relatively brittle, potential $\sigma = 0.4d$, at relatively low twist, before the onset of defects, optimal bundles become anisotropic.  At $\Omega d =0.03$ the cross section becomes weakly anisotropic $\alpha \simeq 1.49$ (compared to the zero twist value of $\alpha = 1.04$) and becomes increasing tapered as twist is raised to $\Omega d = 0.06$ where $\alpha \simeq 2.94$.  For larger twist, optimal bundles return to relatively circular boundary shapes (defined according to $\alpha < 1.5$), and punctuated with the same excess of 5-fold disclinations as the ductile potential states. Thus, for sufficiently brittle inter-filament potentials, anisotropic, defect-free bundles appear intermediate to defect-free cylinders at low-twist and defective cylinders at high twist.

In Fig. \ref{phases}A, we present the morphology phase diagram of  $N=181$ bundles spanned by ductility, $\sigma/d$ (ranging from $0.253 - 1.318$) and reduced twist $\Omega \sqrt{A}$ where $A \equiv N \rho_0^{-1}$ with $\rho_0^{-1} = \sqrt{3} d^2/2$ the area per filament in a uniform hexagonal packing.  Consistent with the above, we observe  that stable defect-free, anisotropic tapes are generally confined to a range of twist intermediate to defect-free and defective cylinders below a critical value of ductility, $(\sigma/d)_c \simeq 1.0$, beyond which the boundary between cylinders and tapes shifts to lower twist with increasingly brittle interactions.  In this range of potentials, the transition from defect-free bundles (either cylinders or tapes) to defective cylinders appears largely independent of ductility, occurring at a critical twist of roughly $(\Omega \sqrt{A})_c \approx 0.84$.  Assuming a circular symmetry, $A= \pi R^2$, this critical twist is consistent with the previous prediction and simulation result for threshold twist at which the relaxation of geometric frustration per 5-fold defect is sufficient to overcome the elastic self-energy for defect formation in isotropic bundles \cite{Grason2010a, Grason2012}.   We note that transitions to higher topological charge packings ($Q=+2,+3,+4$) also appear to be relatively independent of ductility, occurring at roughly the same reduced twist for all values of $\sigma/d$.

\begin{figure}
\centerline{
\includegraphics[width=0.95 \linewidth]{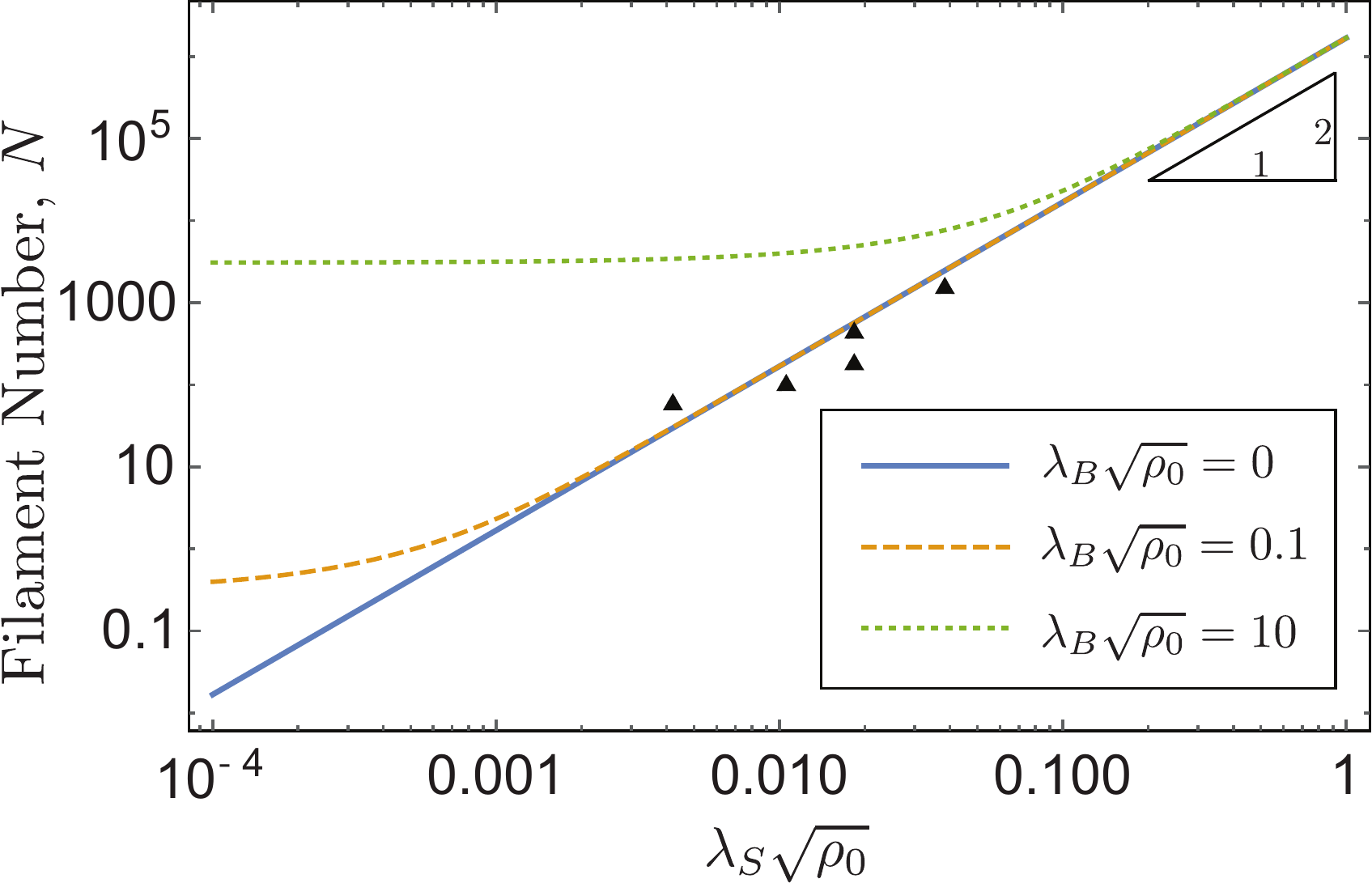}
}
\caption{
Triple point relation between filament number $N = A \rho_0$ and maximum $\lambda_S \sqrt{\rho_0}$ for which tapes are stable. The black triangles indicate results from $N=59,101, 181, 440$ and 1540 simulations where triple point $\lambda_S \sqrt{\rho_0}$ was taken as the highest $\sigma/d$ value at which a defect-free tape structure was found. Curves are predicted from continuum model where the defect-free cylinder/tape boundary (Appendix ~\ref{sec Continuum theory of fixed-area bundles}) crosses the critical threshold for 5-fold defect stability in cylinders, $(\Omega \sqrt{A})_c \approx 0.84$.
}
\label{triple}
\end{figure}

In Fig. \ref{phases}B, we show the morphology phase diagram for a smaller bundle, with $N=59$, which shows the same generic trends as the larger bundle discussed above.  While the threshold reduced twists for stabilizing defective bundles are not observed to change significantly from the $N=181$ case, the range of stable tapes shifts to relatively lower values of $\sigma/d$ (more brittle interactions) with defect-free tapes stable only below $(\sigma/d)_c \simeq 0.48$ for $N=59$.  We observe the same basic trends in the phase diagram (not shown) for bundles at intermediate ($N=101$) and larger ($N=440$ and $N=1540$) filament number, with defect free cylinders for low-twist/ductile interactions, defect-free tapes for intermediate-twist/brittle interactions and defective cylinders at high twist, above $(\Omega \sqrt{A})_c \approx 0.84$.  Moreover, we observe the critical ductility $(\sigma/d)_c$, above which anisotropic cylinders do not form and pass directly from low-twist defect-free cylinders to high-twist defective cylinders, to increase with $N$, as shown in Fig. \ref{triple}.

\begin{figure}
\centerline{
\includegraphics[width=0.9 \linewidth]{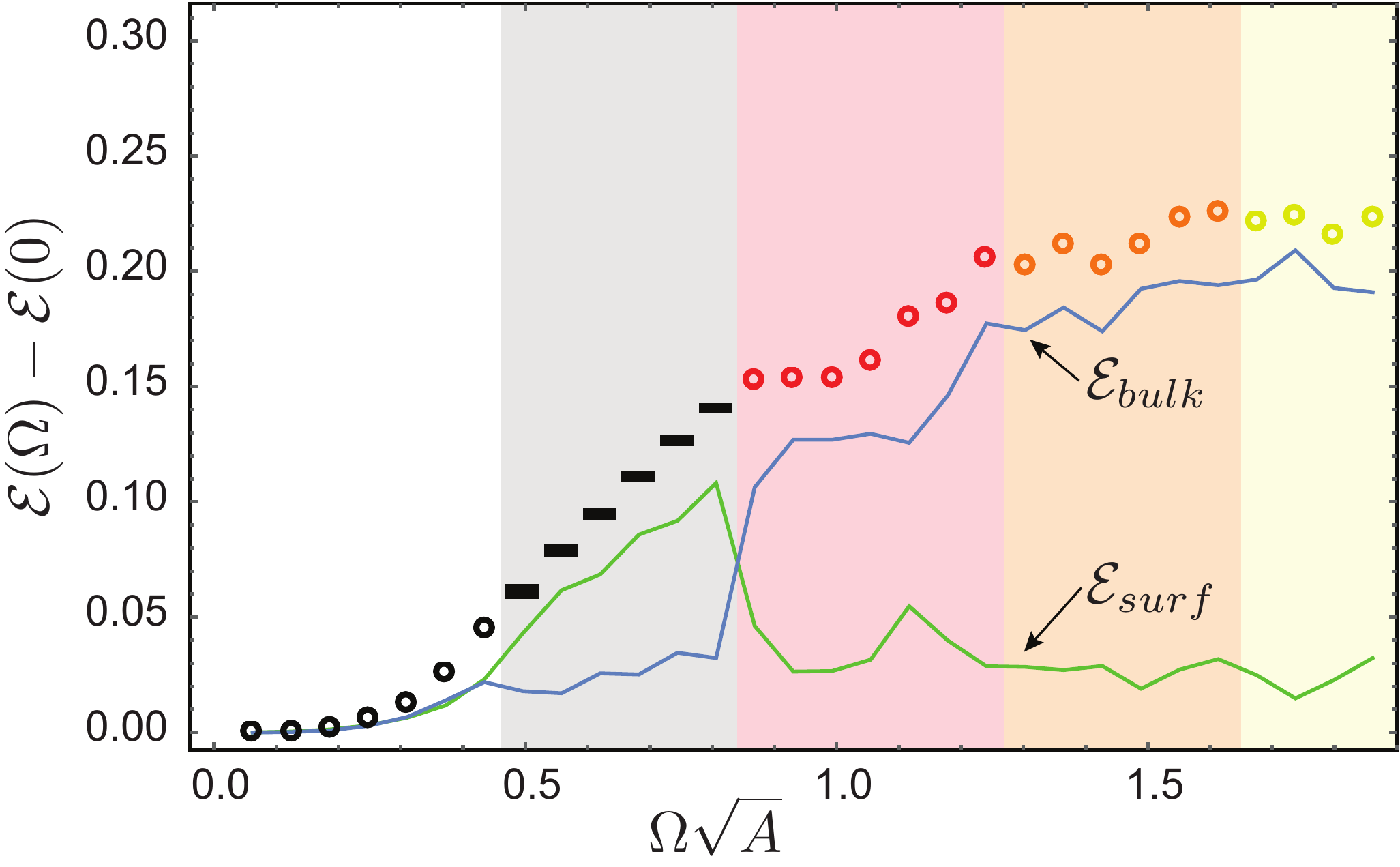}
}
\caption{
Surface, bulk, and total cohesive energy density of simulated ground states (relative to $\Omega=0$ value), for $N=181$ and $\sigma / d = 0.30$. Points show total energy ${\cal E}$, with morphology indicated as in Fig. 5 legend. The green and blue lines show surface (${\cal E}_{surf}$) and bulk (${\cal E}_{bulk}$ filament contributions, where ${\cal E}_{bulk} = {\cal E} - {\cal E}_{surf}$ and ${\cal E}_{surf}$ is the cohesive energy of surface filaments relative to unstrained 6-fold order, i.e. eq. (\ref{Esurf}).  Shaded vertical, regions highlight transitions between stable morphology. }
\label{surf_bulk}
\end{figure}

To understand the observed morphology transitions, in Fig. \ref{surf_bulk} we analyze the energetics of ground-state bundles for a sequence of increasing twist at fixed $\sigma/d=0.3$ and $N=181$ (i.e., sufficiently brittle to form tapes) and compare to continuum theory predictions at fixed $N$ and $B=\lambda_B=0$ (see Appendix \ref{sec Continuum theory of fixed-area bundles}).  Specifically, we analyze the cohesive energy (relative to the $\Omega=0$ state), decomposed in the two pieces, $E_{bulk}$ and $E_{surf}$, the energy of interior and surface filaments, respectively (see Appendix \ref{subs Simulation analysis}).  The energy of defect-free cylinders increases with small twist (both in the bulk and at boundary), in accordance with the elastic penalty due to frustration, $\approx Y A^2 \Omega^4$, from eq. (\ref{felascyl}).  For the fixed-$N$ case, the transition to tapes can be estimated by the point at which the the inter-filament elastic energy of cylindrical bundles grows in excess of the surface energy cost, or $\Sigma \sqrt{A}\gtrsim Y A^2 \Omega^4$, indicating that it is advantageous to create additional surface in order to relax bulk frustration.  A more careful analysis (see Appendix \ref{sec Continuum theory of fixed-area bundles}) gives a reduced twist for the cylinder-tape transition $(\Omega \sqrt{A})_{cyl/tape} \simeq 5 (\lambda_S/\sqrt{A} )^{1/4} \propto (\sigma/d)^{1/2}N^{-1/8}$, where we used $Y\approx \epsilon/\sigma^2$, $\Sigma \approx \epsilon/d$ and hence, $\lambda_S \approx \sigma^2/d$.  These values of reduced twist are shown as blue curves in the phase diagrams in Fig. \ref{phases}, and agree well with the defect-free cylinder-to-tape transition twist, and more significantly, its tendency to {\it decrease} with $N$ and increase with $\sigma/d$, depending on the relative cost of inter-filament strain in comparison to bundle surface.  

\begin{figure}
\centerline{
\includegraphics[width=0.9 \linewidth]{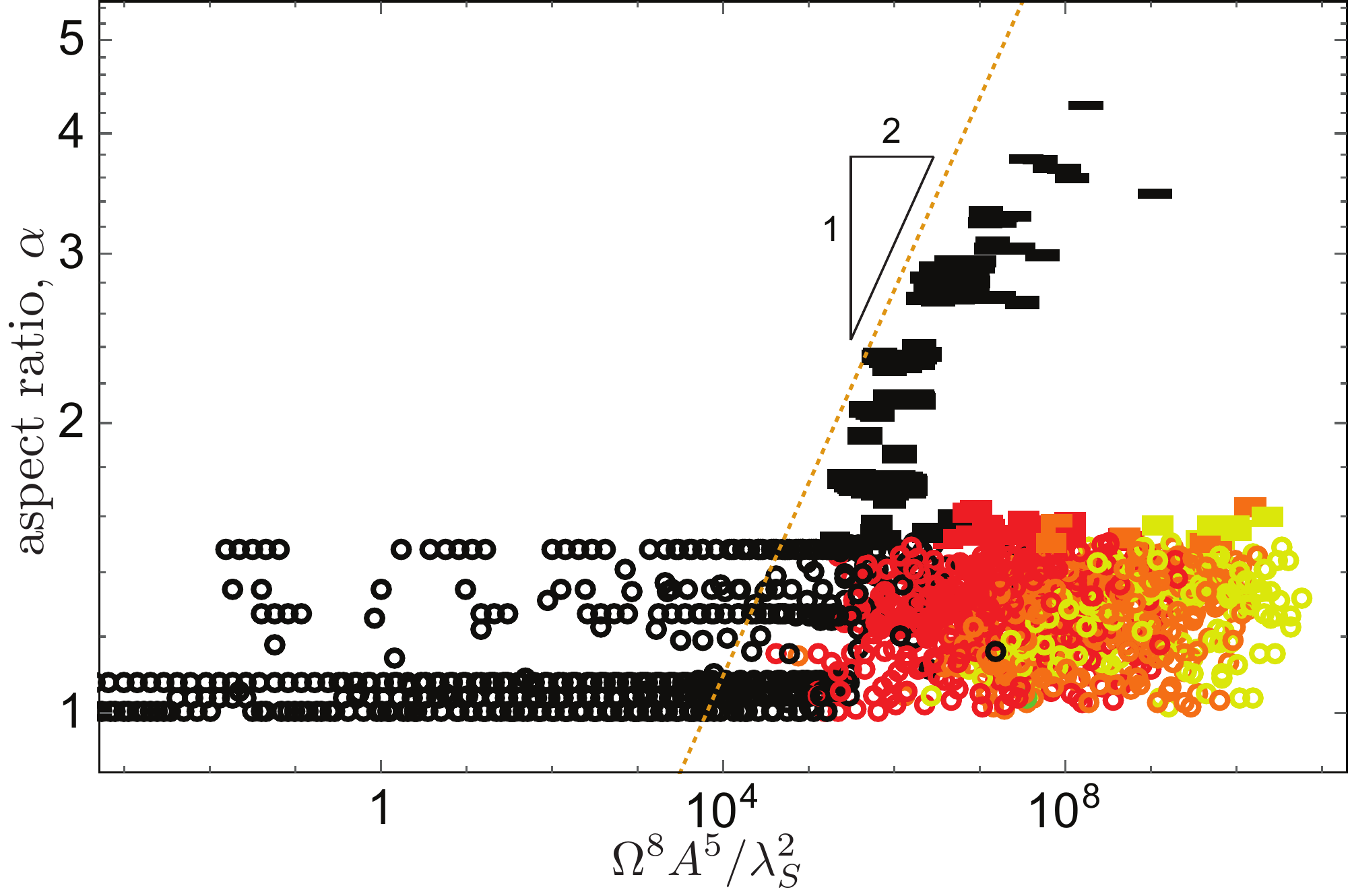}
}
\caption{
Optimal aspect ratio, $\alpha$, for simulation bundle ground states plotted versus a dimensionless ratio of predicted bulk-frustration to surface energy penalty for cylinders, $A^{5/2} \Omega^4/\lambda_S$, for values of $N=59, 101$ and $181$.  The dashed line shows large-$\alpha$ power law predicted from fixed-$N$ continuum model, with morphology indicated as in Fig. 5 legend.  Note that defective morphologies fall below $\alpha \simeq 1.5$, the threshold we use to discriminate tapes from cylinders.  }
\label{aspectscale}
\end{figure}

As shown in Fig. \ref{surf_bulk}, at an intermediate range of reduced twist  $(\Omega \sqrt{A})_{cyl/tape}<\Omega \sqrt{A} < (\Omega \sqrt{A})_c$, anisotropic bundle shapes become favorable.  From the continuum model (see Appendix \ref{sec Continuum theory of fixed-area bundles}), the equilibrium aspect ratio (for $\alpha \gg 1$) is determined by a balance of excess surface energy, growing with narrow dimension $w = \sqrt{A \alpha}$ as $\approx \Sigma \sqrt{A \alpha}$, and inter-filament strain growing with thin dimension $t = \sqrt{A/\alpha}$ as $\approx Y \Omega^4 (A / \alpha)^2$, yielding an equilibrium $\alpha_* \approx A (\Omega^4/\lambda_S)^{2/5}$ that increases with filament number, twist and brittleness of inter-filament bonds.  This power-law trend for tape aspect ratio is compared to simulated bundles in Fig. \ref{aspectscale}.  Because equilibrium anisotropy increases with twist $\Omega$, The surface energy $E_{surf}$ of tapes increases rapidly with twist, while $E_{bulk}$ remains relatively constant with $\Omega$ due to the mitigation of inter-filament strain by the narrow bundle thickness, as shown for stable tapes at intermediate-twist $0.46 < \Omega \sqrt{A} < 0.84$ in Fig. \ref{surf_bulk}.

At high twist (above $(\Omega \sqrt{A})_c \simeq 0.84$), defective cylinders become stable, leading to a drop of surface energy due to the fewer number of boundary filaments in the circular cross section.  For high twist, $E_{bulk} > E_{surf}$ due to the straining of inter-filament bonds by the combined effect of frustration and topological defects.  As reported  previously in ref. \cite{Bruss2013} the steady increase of $Q$ with $\Omega$ leads to a nominal plateau of energy density at high twist as interior defects effectively mediate the cost of increased frustration by twist.  This generic flattening of the bulk frustration energy due to one or more excess 5-fold disclinations is responsible for the generic crossover from stable intermediate twist tapes, with energies that grow for modestly with twist ($f_{tape} \sim \Omega^{4/5}$ as compared to $f_{cyl} \sim \Omega^4$ for cylinders), to defective cylinders, which approach a relatively constant, if somewhat cuspy energy density due to the fixed circular boundary and neutralizing effect of 5-fold disclinations.

\begin{figure}
\centerline{
\includegraphics[width=0.9 \linewidth]{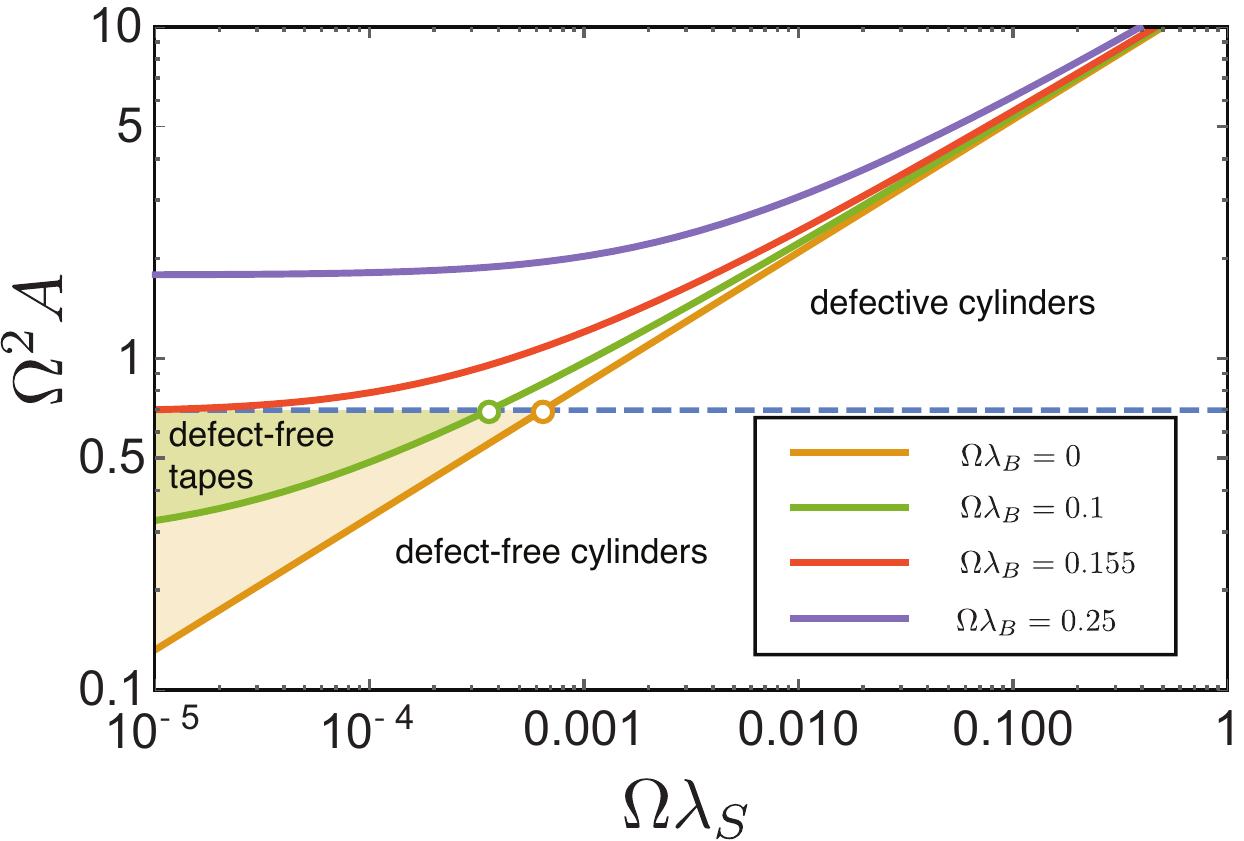} 
}
\caption{ Continuum theory (fixed-$N$/fixed-$\Omega$) phase diagrams for morphology selection of twisted bundles for a range of values of $\Omega \lambda_B$.  Solid lines are predicted transition lines between defect-free cylinders and tapes, and dashed line is threshold stability for 5-disclinations in cylinders.  Equilibrium tapes are predicted in the shaded regions between the solid boundaries and dashed lines, with triple points indicated as empty circles.  For $\Omega \lambda_B \geq 0.155$, tape morphologies are not stable.}
\label{contphase}
\end{figure}

We use the predictions of the continuum model to present an overview of the fixed-$N$ morphology selection in Fig.~\ref{contphase}, here as a phase diagram spanned by the axes $\Omega^2 A$ and $\Omega \lambda_S$.   The intersection of the cylinder-tape boundary $\Omega^2 A \simeq 13.2 \Omega \lambda_S$ with the critical condition for stabilizing 5-fold defects in cylinders $\Omega^2 A \simeq 0.71$ yields the triple point where defect-free cylinder, defect-free tape and defective cylinder phases meet, giving also the prediction for the dependence of the critical ductility $(\sigma/d)_c$ on $N$, shown in Fig.~\ref{triple}.   Here, we also show the effect of non-zero bending stiffness or $\Omega \lambda_B \neq 0$.  The effect of bending stiffness in this morphology phase diagram is to penalize tape morphologies (which possess a larger second moment of cross-section area $\langle r^2 \rangle$) relative to cylinders, shifting the cylinder-tape boundary to larger values of $\Omega^2 A$ as shown in Fig.~\ref{contphase}, particularly at low values of $\Omega \lambda_S$ where brittle interactions otherwise favor very-wide tapes.  This analysis shows that above a critical stiffness value, for $(\Omega \lambda_B)_{max} \geq 0.155$, the window of intermediate twist tapes is shifted above the threshold for stable defects (which are insensitive to bending stiffness) and we therefore expect tape-like bundles to be preempted entirely by defective cylinders when filaments exceed this critical stiffness $B_{max} \approx \epsilon \Omega^{-2} (d/\sigma)^2$.

\section{Conclusions}

In this paper we investigate two models of morphology competition for self-assembled chiral filament bundles.  The first considers the ability to ``escape" frustration by untwisting chiral inter-filament skew and its effect on the otherwise self-limiting assembly of bundles.  We find that finite stiffness of inter-filament chiral interactions, $K$, puts an upper limit on the (cohesive) surface energy $\Sigma_{max} \sim K^\nu$ at which finite, frustration-limiting bundles form where $\nu =3/2$ for $K \ll K_c$ and $\nu =5/4$ for $K \gg K_c$ where $K_c = \Omega_0^2 (\rho_0 B)^2/Y$.  When chiral interactions are sufficiently stiff ($K \gtrsim 100 K_c$) equilibrium assembly includes  windows of both cylindrical (small $\Sigma$) and tape-like bundles (intermediate $\Sigma$) as well as bulk, untwisted assembly (large $\Sigma$), while in the opposite limit of ``soft" chiral interactions, the window of tapes narrows dramatically and ultimately vanishes, leading to direct transition from twisted cylinders to bulk assembly for even weakly cohesive bundles.  The second model investigates the interplay between competing morphological responses to inter-filament frustration:  bulk defects vs. boundary anisotropy.  We found that when inter-filament bonds are sufficiently brittle, tape-like bundles are stable as an intermediate-twist phase between defect-free cylinder (low twist) and defective cylinder (high twist) morphologies, and that the selection of the energetically favorable morphology was controlled by the relative cost of straining inter-filament bonds in the bulk (as required to form large isotropic domains {\it and} topological defects) compared to the cohesive cost of creating excess free bundle surface, which has the critical effect of lowering frustration costs through low-energy boundary deformation.  

The results demonstrate a surprising and generally overlooked properties of geometrically-frustrated assemblies \cite{Grason2016}.  Preferred chiral ordering makes uniform order, both intra- and inter-filament, impossible, leading to a competing set of heterogeneous, and morphological distinct, local free energy minimal.  Thus, a proper understanding of even the equilibrium behavior of these systems is a far greater challenge than canonical, unfrustrated systems, requiring an analysis of all classes of metastable equilibrium, each of which is governed by a distinct set of thermodynamic or microscopic parameters describing assembly.  In the context of the present work, an outstanding challenge still remains to compare and understand the competition of {\it all four classes} of morphological responses at once:  self-limiting width, untwisting of chiral interactions, formation anisotropic boundary shape or topological defect in the interior.  

Furthermore, a much broader challenge will remain to understand how the principles of morphology selection analyzed here also controls non-equilibrium assembly.  For example, in \cite{Hall2016} we analyzed the tape-like and cylindrical morphologies of twisted amyloid protein bundles in terms of the equilibrium (fixed twist) model of domain shape selection, and we find that tapes form at a degree of bundle twist, where based on the present study, we would expect equilibrium morphologies to relax frustration via one or more 5-fold interior defects while retaining a cylindrical shape. The observation of tapes in this regime suggests a kinetically limited assembly pathway, where the bundles are kinetically free to relax frustration upon reaching a critical size through anisotropic domain growth, but may be subject to very large barriers that inhibit the global reorganization of defect-free tapes into defective cylinders.  A systematic study of the thermodynamically and kinetically accessible pathways {\it between} distinct morphologies of chiral filament bundles, or geometrically frustrated assemblies more broadly, remains a critical and unresolved aspect of the anomalous behavior of these systems.

\begin{acknowledgments}
The authors are grateful to acknowledge E. Matsumoto, G. Schr{\"o}der-Turk and I. Bruss for stimulating discussions, as well as D. Atkinson and Q. Meng for helpful comments on this manuscript.  This research was supported by the NSF through grant no. DMR 16-08862 and by the Donors of the American Chemical Society Petroleum Research Fund through ACS PRF grant no. 54513-ND. Numerical simulations were performed on the UMass Shared Cluster at the Massachusetts Green High Performance Computing Center.
\end{acknowledgments}


\appendix

\section{Continuum theory of fixed-area bundles} \label{sec Continuum theory of fixed-area bundles}

Here, we describe the continuum model prediction for coexistence between cylindrical and tape morphologies, restricted to fixed  twist rate $\Omega$ and fixed area $A$, equivalent to fixed filament number, $N \simeq A \rho_0$.  In this case, cylinder radius $R$ and tape width $w$ and thickness $t$ satistfy $A=\pi R^2 = w t$.  Defining dimensionless area $\bar{  A} \equiv \Omega^2 A$, elasto-cohesive length $\bar{ \lambda} _S \equiv \Omega \lambda_S$, and bend penetration depth $\bar{ \lambda} _B \equiv \Omega \lambda_B$, we rewrite  eqs. \ref{felascyl} and \ref{felastape} (neglecting the contribution from twist elasticity for fixed-twist ensemble and re-casting energy density $f \equiv F / YV$),
\begin{equation} \label{eq fixed cyl}
f_{\rm cyl} = \frac{2 \pi^{1/2} ~ \bar{ \lambda} _S}{\bar{ A}^{1/2}} + \frac{\bar{ \lambda}_B ^2}{4 \pi} \bar{ A} + \frac{3}{128 \pi^2} \bar{ A}^2
\end{equation} and \begin{multline} \label{eq fixed tape}
f_{\rm tape}\simeq \frac{2 ~ \bar{ \lambda}_S}{\bar{ A}^{1/2}} \Big( \sqrt{\alpha} +\frac{1}{\sqrt{\alpha}}\Big)+\frac{\bar{ \lambda}_B^2 }{24}\bar{ A}(\alpha + \frac{1}{\alpha}) \\ \\ +\frac{1}{160}\bar{ A}^2  \alpha^{-2} .
\end{multline} 
The equilibrium value of aspect ratio $\alpha*$ is determined by the condition
\begin{equation}\label{eq: eqm alpha}
\frac{\partial f_{\rm tape}}{\partial \alpha} \bigg|_{\alpha*}= 0
\end{equation}
where we consider only the stable solutions where aspect ratio $\alpha* > 1$, since the inter-filament elastic energy fails to hold for $\alpha \lesssim 1$.  For example, if we consider the limit of $\alpha \gg 1$ and $\lambda_B =0$ (as discussed in the text), we may neglect the bending terms and the surface contribution from the narrow ends of the tape section ($\propto \alpha ^{-1/2}\ll1$), relative to the large contribution from the wide dimension ($\propto \alpha ^{1/2}\gg1$).  From this limit of \ref{eq: eqm alpha} we find equilibrium tape dimensions and energy,
\begin{equation} 
\alpha^* = \frac{\bar{ A}}{(80~ \bar{\lambda}_S)^{2/5}}, ~
f_{\rm tape} (\alpha^*) = \frac{5^{4/5}}{2^{9/5}} \bar{\lambda}_S^{4/5}
\end{equation}
Coexistence is determined by the solution to $f_{\rm cyl} (\alpha*  = f_{\rm tape}$.

Exact numerical solution of equilibrium conditions for $\lambda_B=0$ gives coexistence between cylinders and tapes along the line
\begin{equation}
\Omega^2 A \simeq 13.2 (\Omega \lambda_S )^{2/5}.
\end{equation}.

With increasing bending cost, tapes become less stable due to the distribution of filaments to larger radial distances from the twist axis, and therefore, higher curvature. Notably, for $\Omega \lambda_B > 0.155$, tapes are unstable for all $\Omega^2 A$ below the threshold $2 \pi / 9$ which defines coexistence of the defect-free cylindrical morphology with defective cylindrical morphology with (+1) topological charge (equivalent to $(\Omega R)_c = \sqrt{2/9}$ derived in \cite{Grason2010a}). The triple point where defective structures coexist with both defect-free tapes and cylinders, is found according to the above conditions for coexistence, subject to the constraint $(\Omega^2 A)_c = 2 \pi / 9\simeq 0.7 $. 

\section{Simulation Methods}

\subsection{Inter-filament potential}\label{subs Discrete filament model}

Here we describe formulas used to model the inter-filament cohesive interactions.  Similar descriptions appeared previously in \cite{Bruss2012a, Bruss2013, Hall2016}.

We assume the form of the pairwise interaction potential defined by $u(\Delta)$ and $\Delta E$ in eqs. \ref{uDelta} and \ref{deltaE} derives from the superposition of pair-wise interaction $V(r)$ between material points along two filaments, as described before in \cite{Bruss2013}. Parameterizing centerlines of $i$ and $j$ filaments by $\Xv_i(s_i)$ and $\Xv_j(s_j)$ ($s_i$ and $s_j$ are arc length coordinates), then the total energy is simply
\begin{equation}
 \Delta E_{ij} = \int ds_i \int ds_j ~ V(|\Xv_i(s_i) - \Xv_j(s_j)|)
\end{equation}
Taking $\Xv_i(s_i)$ as the outer filament (such that $\rho_i \geq \rho_j$), we define $\bD_{ij}$ as the separation that connects $\Xv_i(s_i)$ with the closest point on $i$ at $\Xv_j\big(s_j^*(s_i) \big)$.  That is  $ \Xv_j(s_j^*(s_i)) -\Xv_i(s_i) = \bD_{ij}$, such that $\frac{d}{d s_j}|\Xv_i(s_i) - \Xv_j(s_j)|^2\big|_{s_j^*(s_i)}=0$.  Defining $\delta s_j \equiv s_j - s_j^*(s_i)$ as the arc-distance from the closest contact on $j$ to $i$, we have
\begin{equation}
|\Xv_i(s_i) - \Xv_j(s_j)|^2 = |\bD_{ij}|^2 + (\delta s_j)^2(1 + \kappa_j \bD_{ij} \cdot {\bf N}_j) + O[(\delta s_j)^3]
\end{equation}
where ${\bf \Delta}_{ij} \cdot {\bf N}_j$ is the projection of the inter-filament separation along the normal direction of the inner filament. For sufficiently short-range potential $V(r)$ with respect to filament curvature and constant filament curvatures:
\begin{equation}
\int ds_j V(|\Xv_i(s_i) - \Xv_j(s_j)|) = \frac{\int du V(\sqrt{|\bD_{ij}|^2 + u^2})}{\sqrt{1 + \kappa_j \bD_{ij} \cdot {\bf N}_j}} 
\end{equation}
Defining 
\begin{equation}
u(\Delta) \equiv \int_{-\infty}^{+\infty} du~V(\sqrt{\Delta^2 + u^2}),
\end{equation} and for outer filament length $\ell_i$ :
\begin{equation}
d E_{ij} = \frac{u(|\bD_{ij}|) }{\sqrt{1 + \kappa_j \bD_{ij} \cdot {\bf N}_j}} d s_i .
\end{equation}
Thus, the total energy depends on the length of the outer filament and the second-order deflection of $j$ near to close contact with $i$, accounting for the factor of $(1 + \kappa_j \bD_{ij} \cdot {\bf N}_j)^{-1/2}$.  Taking $\Delta \ell_i$ as the discrete approximation for $ds_i$ we arrive at the formula for $ \Delta E_{ij}$ used in eq. (\ref{deltaE}).

When the interaction has the form of a Lennard-Jones potential, $V(r) = V_0 [(\sigma_0 / r)^{12} - 2(\sigma_0 / r)^6]$, then $u(\Delta) = \frac{\epsilon}{6}[5(\sigma / \Delta]^{11} - 11(\sigma / \Delta)^5]$ with preferred filament spacing $d = \sigma=0.947 \sigma_0$ and minimum energy $\epsilon = 1.686 V_0 \sigma_0$ \cite{Bruss2013}.  Hence the range (and curvature) of the cohesive potential is locked to the equilibrium spacing for this particular model of cohesive ``threads".  Here, we use a broader class of inter-filament potentials, $u(\Delta)$, that allow for independent adjustment of equilibrium spacing, $d$,  and range, $\sigma$ of attractive well:
\begin{equation}
u(\Delta) = \frac{\epsilon}{6}\Big[\frac{ 5 \sigma^{11}}{(\Delta + \sigma - d)^{11} }- \frac{ 11 \sigma^{5} }{(\Delta + \sigma-d)^5 } \Big]
\end{equation} 
At the minimum $\Delta = d$, the curvature, or stiffness, of the potential is a function of $\sigma$ only,
\begin{equation}
\frac{d^2 u(\Delta)}{d \Delta^2} \bigg|_{\Delta=d} = \frac{55 \epsilon }{ \sigma^2}
\end{equation} 
The above form of this potential at the minumum controls the coarse-grained modulus of the inter-filament array (see \cite{Bruss2013, Hall2016}), $Y \simeq 110 \epsilon /(\sqrt{3} \sigma^2 )$, while the filament diameter controls the surface energy $\Sigma \simeq \epsilon / d$, allowing for a broadly tunable range of elasto-cohesive length $\lambda_S = \Sigma/Y$

The inter-filament spacing $\Delta$ is computed according to an approximation previously established \cite{Bruss2012a}. In short, for filaments with centerlines intersecting the cross section at positions ${\xv}_i$ and ${\xv}_j$, the exact expression for interfilament distance comes from minimization with respect to the vertical component of separation $z$ of the following:
\begin{multline}\label{eqDOCA}
|{\bf \Delta}_{ij}(z)|^2 = |{\bf x}_i|^2 + |{\bf x}_j|^2 - 2 {\bf x}_i  \cdot {\bf x}_j \cos(\Omega z)  \\ + 2 |{\bf x}_i \times {\bf x}_j| \sin(\Omega z) + z^2
\end{multline}
To approximate the contact separation $z_*$, we use the following interpolation,
\begin{equation}\label{eqDOCA2}
\text{tan}(\Omega z_*) \approx  \frac{\Omega^2 |{\bf x}_i \times {\bf x}_j| } { 1 + \Omega^2 {\bf x}_i  \cdot {\bf x}_j}.
\end{equation}
This form interpolates between the exact solutions to $\frac{d}{dz}|{\bf \Delta}_{ij}(z)|^2=0$, in the limits of small and large $\Omega^2 \rho_i \rho_j$ and has been shown to be sufficiently accurate for potentials whose range is smaller that their thickness, a limit well satisfied by simulations here.

\subsection{Simulation Details}\label{subs Simulation Details}

\begin{figure}
\centerline{
\includegraphics[width=0.9 \linewidth]{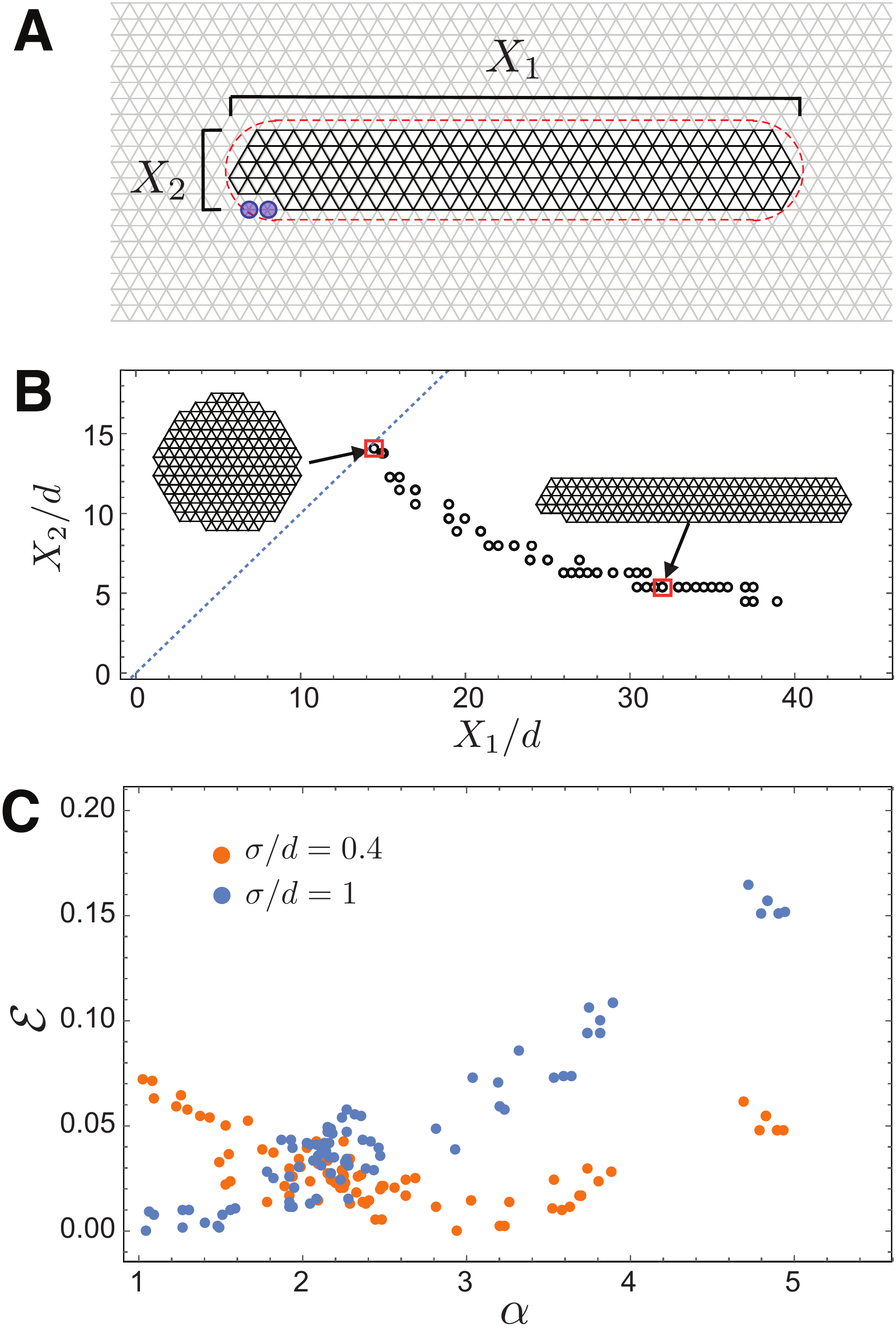}
}
\caption{
In (A), starting configurations are sampled with filament centerline positions arranged on an hexagonal lattice. Structures of varying aspect ratio are taken by selecting points within a stadium boundary with varying dimensions. Lattice vertices circled blue correspond to ``pruned" filament positions removed to maintain target $N$.  In (B), the dimensions of the lattice configurations sampled for $N=181$ filaments are shown, corresponding to sampling with stadium boundary aspect ratios $\alpha =1 - 7.5$, with example configurations are shown as highlighted. In (C), the energy density of relaxed lattice configurations are shown at twist $\Omega \sqrt{N} = 0.80$, for two values of ductility $\sigma / d$.  For brittle case, $\sigma / d = 0.4$, $\alpha \simeq 3$ tapes are favored, while for the relative ductile case, $\sigma / d = 1.0$, roughly cylindrical bundles ($\alpha \simeq 1$) are optimal.}
\label{simdet}
\end{figure}

{\it Method A, Lattice Sampling} - Initial configurations were generated with filament positions on a triangular lattice with spacing $d$, at points within a stadium perimeter (connecting two half-circles with straight lines as in Figure 1A). The stadium aspect ratios were taken ranging from 1.0 to 7.5 in increments of 0.2, and taken with center on a lattice point, lattice edge and the center of of a triangular plaquette. For aspect ratio larger than 1, the longer dimension (tape width) was oriented along the close-packed lattice direction. Stadium dimensions were chosen to be the smallest so that each choice of aspect ratio and stadium center enclosed the desired number of filaments. When this produced a structure with excess filaments (i.e. close to, but more than predetermined $N$), additional points were omitted from the last close-packed row, starting from the end. This filament ``pruning" is illustrated in Fig. \ref{simdet}A, where two outer points are omitted to produce the structure with 181 filaments. Fig. \ref{simdet}B shows the dimensions, and two limiting example structures, of lattice initialized configurations sampled for $N = 181$ filaments. Corresponding relaxed energies for each lattice structure sampled are shown in Fig. \ref{simdet}c for a twisted bundles at two sample values of ductility $\sigma / d$ (brittle/ductile). 

\begin{figure}
\centerline{
\includegraphics[width=0.9 \linewidth]{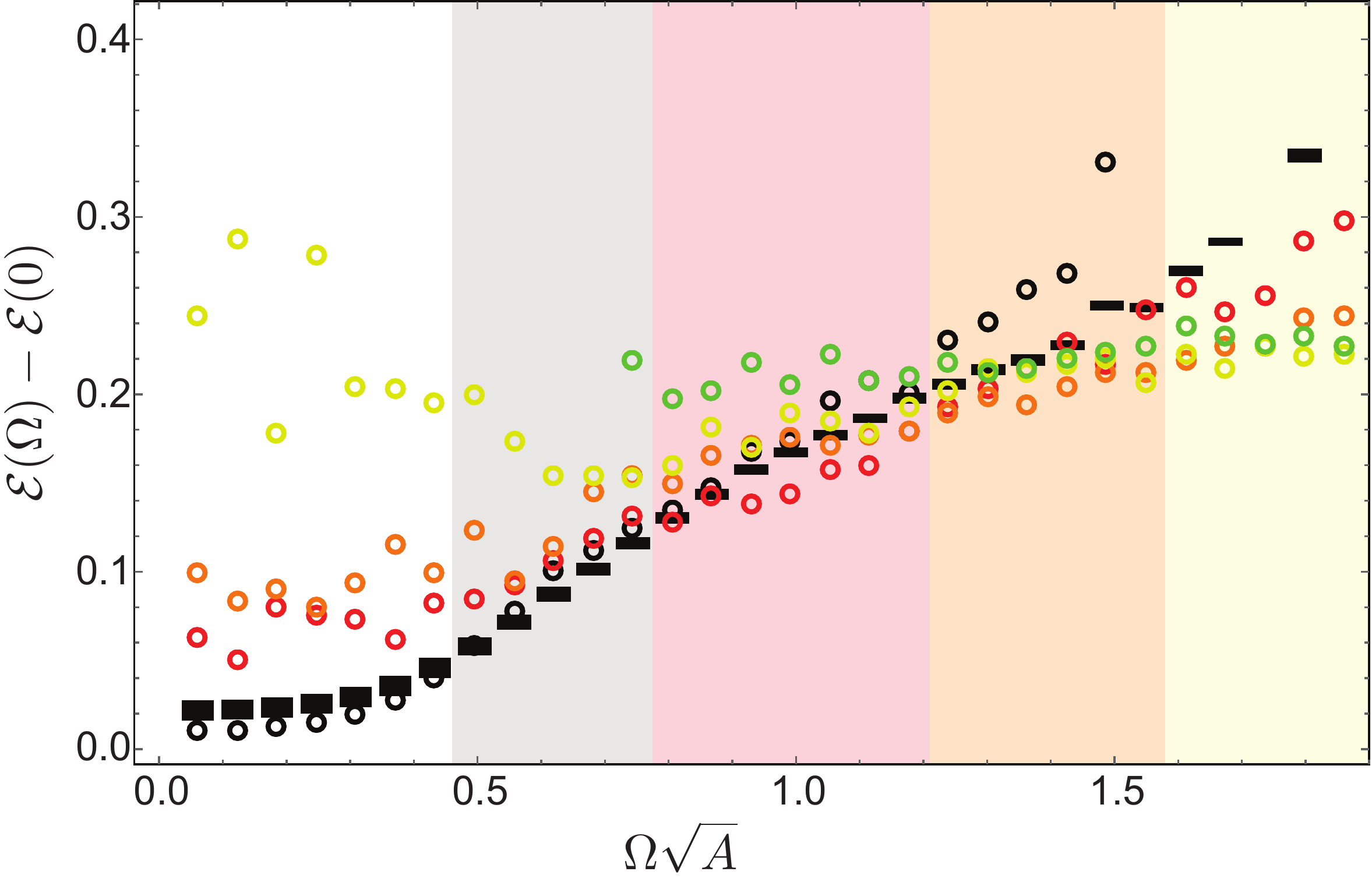}
}
\caption{
Lowest-energy states of competing morphologies, for $N=181$, $\sigma / d = 0.40$, plotted according to cohesive energy density (relative to $\Omega =0$ value). Here, we show lowest energy obtained with each morphological class, with morphology indicated as in Fig. 5 legend: defect-free cylinders or tapes and defective cylinders for each non-zero $Q$ value.  Shaded vertical regions highlight transitions between stable morphology. }
\label{evsQ}
\end{figure}

{\it Method B, Random Sampling} - Random initial configurations were generated by uniformly distributing filament positions within a circular boundary of radius  $0.5 \sqrt{N} d$, such that the starting configuration always had slight compression. In Fig. \ref{evsQ}, we show how these randomly seeded configurations lead to competing energy minima with a range of $Q$ for a given bundle parameter values. For the purposes of our present study, we chose to exhaustively sample the defective configurations for $N=59$ and $N=181$, and sample defect configurations in $N=101$ and $N=440$ bundles more sparsely, only to confirm the transition  from defect-free cylinders/tapes to defective packings at the critical twist.	 Accordingly, the number of randomly-initial configurations, $n_{init}$ per set of bundle parameters ($\Omega$ and $\sigma/d$) was varied with $N$: $n_{init}=6000$ for $N=59$; $n_{init}=3000$ for $N=101$; $n_{init}=60000$ for $N = 181$; and $n_{init} = 3000$ for $N=440$.  For $N=1540$, no attempt was made to sample random initial structures, due to the prohibitive computational cost for bundles at this size.  

For each value of filament number $N$, simulations were run with twist $\Omega \sqrt{N}$ ranging from 0 to 2.0 with 30 equally incremented values. The range of ductility $\sigma / d$ varied with filament number $N$, and values of $\sigma / d$ were spaced logarithmically with successive ratio 1.096:
\begin{center}
\begin{ruledtabular}
\begin{tabular}{|r | r | r|}
$\#$ filaments $N$ &  Ductility $\sigma / d$ & $\#$ random samples \\
\hline
 59 & 0.253 - 1.318    & 6 000\\ \hline
 101 &  0.253 - 1.318  & 3000\\ \hline
 181 & 0.253 - 1.318   & 60 000\\ \hline
 440 & 0.253 - 1.738  & 3000\\ \hline
 1540 &  0.253 - 1.738   & N/A
\end{tabular}
\end{ruledtabular}
\end{center}

{\it Structure Relaxation} - From each set of initial filament positions, a local equilibrium was found by steepest descent according to the defined energy density, stopping when the largest component of the gradient (parameterized by filament coordinates) had magnitude less than $10^{-4} \epsilon /d$. In areas of low twist, the best structures were generated by slight relaxation of configurations generated from a triangular lattice. Larger rearrangements were required in regions of high twist, and randomly generated starting configurations consistently produced more favorable structures. In order to allow larger rearrangements even with brittle systems and to remove overlaps with filament ``cores" ($\Delta < d-\sigma$), initial structures were first (before steepest descent) preconditioned by relaxation via steepest descent according to a soft repulsive potential,
\begin{equation}
U_{pre} = \sum_{i,j}^N   (1-|\Delta_{ij}|/d)^2 \Theta(d-|\Delta_{ij}|)  \end{equation}

\subsection{Simulation analysis} \label{subs Simulation analysis}

{\it Morphology Characterization} - The lowest-energy structure at each point on the phase diagram was analyzed, taking both the aspect ratio and net topological charge of the interior.  The aspect ratio was taken from the lateral dimension of the minimum-area rectangle that encloses the cross section.   The minimal-area rectangle is found from considering all rectangles enclosing the structure's convex hull where one edge of the rectangle must coincide with an edge of the convex hull.

The net topological charge was computed by Delaunay triangulation of the transformed cross-sectional positions.  As the true distances between filaments are properly encoded in the dual metric surface, we triangulate based on a conformal map of filament positions from the dual surface to the 2D plane .  This map is performed by radial dilation of filament position, $\rho' = \frac{\rho}{1+\sqrt{1+(\Omega \rho)^2}}e^{\sqrt{1+(\Omega \rho)^2}}$ for a filament with cross-sectional radial position $\rho$. This mapping results in the angles between filament separations reflecting the true inter-filament bond angles when measuring filament distances according to the geodesic distance on the dual surface \cite{Bruss2012a}. The conformal map allows regular Delaunay triangulation, subdividing the convex hull of the packing into triangles connecting filament positions, so that the minimum inter bond-angle of the triangulation is maximized. 

The perimeter of the bundle was constructed by starting with the convex hull, then successively removing exterior bonds with large distance of closest approach, $|(\Delta_{ij}/d)| > 1.4$ and adding the interior point that shared a triangular plaquette according to the Delaunay triangulation. The net topological charge of the interior was then taken by totaling the excess coordination for each filament not belonging to the set of boundary filaments.

{\it Surface and Bulk Cohesive Energies} - Here we decompose the cohesive contribution from pair-wise interactions,
\begin{equation}
\Delta E_{\text{coh}} = \sum_{\rho_j \leq \rho_i} \frac{u(|\Delta_{ij}|) \Delta \ell_i}{\sqrt{1 + \kappa_j {\bf \Delta}_{ij} \cdot {\bf N}_j}} ,
\end{equation}
into energies associated with {\it surface formation} and {\it bulk inter-filament deformation}. The surface energy is taken to be the deficit of exterior filament contact from a reference state with total cohesive energy $3 \epsilon$. The surface energy is then 
\begin{equation}
\label{Esurf}
\Delta E_{\text{surf}} =  \sum_{i \in \text{exterior}} \bigg[ 3 \epsilon - \Big(\frac{1}{2} \sum_{j \leq N} \frac{u(\Delta_{ij}) }{\sqrt{1 + \kappa_j {\bf \Delta}_{ij} \cdot {\bf N}_j}} \Big)  \bigg] \Delta \ell_i
\end{equation}
The bulk energy is taken as the difference, $ \Delta E_{\text{bulk}} = \Delta E_{\text{coh}} - \Delta E_{\text{surf}}$.


\begin{thebibliography}{10}
\expandafter\ifx\csname urlstyle\endcsname\relax
  \providecommand{\doi}[1]{doi:\discretionary{}{}{}#1}\else
  \providecommand{\doi}{doi:\discretionary{}{}{}\begingroup
  \urlstyle{rm}\Url}\fi

\bibitem{Harris1999}
Harris, A.~B., Kamien, R.~D. \& Lubensky, T.~C., 1999 {Molecular chirality and
  chiral parameters}.
\newblock \emph{Reviews of Modern Physics} \textbf{71}, 1745--1757.
\newblock ISSN 0034-6861.
\newblock (\doi{10.1103/RevModPhys.71.1745}).

\bibitem{Straley1976}
Straley, J.~P., 1976 {Theory of piezoelectricity in nematic liquid crystals,
  and of the cholesteric ordering}.
\newblock \emph{Phys. Rev. A} \textbf{14}, 1835--1841.
\newblock (\doi{10.1103/PhysRevA.14.1835}).

\bibitem{Goodby1991}
Goodby, J.~W., 1991 {Chirality in liquid crystals}.
\newblock \emph{Journal of Materials Chemistry} \textbf{1}, 307--318.
\newblock ISSN 0959-9428.
\newblock (\doi{10.1039/JM9910100307}).

\bibitem{Selinger2001}
Selinger, J.~V., Spector, M.~S. \& Schnur, J.~M., 2001 {Theory of
  Self-Assembled Tubules and Helical Ribbons}.
\newblock \emph{The Journal of Physical Chemistry B} \textbf{105}, 7157--7169.
\newblock ISSN 1520-6106.
\newblock (\doi{10.1021/jp010452d}).

\bibitem{Ghafouri2005}
Ghafouri, R. \& Bruinsma, R., 2005 {Helicoid to Spiral Ribbon Transition}.
\newblock \emph{Physical Review Letters} \textbf{94}, 138101.
\newblock ISSN 0031-9007.
\newblock (\doi{10.1103/PhysRevLett.94.138101}).

\bibitem{Armon2014}
Armon, S., Aharoni, H., Moshe, M. \& Sharon, E., 2014 {Shape selection in
  chiral ribbons: from seed pods to supramolecular assemblies}.
\newblock \emph{Soft Matter} \textbf{10}, 2733--2740.
\newblock (\doi{10.1039/C3SM52313F}).

\bibitem{neville1993biology}
Neville, A.~C., 1993 \emph{{Biology of Fibrous Composites: Development Beyond
  the Cell Membrane}}.
\newblock Cambridge University Press.
\newblock ISBN 9780521410519.

\bibitem{Bouligand2008a}
Bouligand, Y., 2008 {Liquid crystals and biological morphogenesis: Ancient and
  new questions}.
\newblock \emph{Comptes Rendus Chimie} \textbf{11}, 281--296.
\newblock ISSN 16310748.
\newblock (\doi{10.1016/j.crci.2007.10.001}).

\bibitem{Sharma2009}
Sharma, V., Crne, M., Park, J.~O. \& Srinivasarao, M., 2009 {Structural Origin
  of Circularly Polarized Iridescence in Jeweled Beetles}.
\newblock \emph{Science} \textbf{325}, 449--451.
\newblock ISSN 0036-8075.
\newblock (\doi{10.1126/science.1172051}).

\bibitem{Brunsveld2001}
Brunsveld, L., Folmer, B. J.~B., Meijer, E.~W. \& Sijbesma, R.~P., 2001
  {Supramolecular Polymers}.
\newblock \emph{Chemical Reviews} \textbf{101}, 4071--4098.
\newblock ISSN 0009-2665.
\newblock (\doi{10.1021/cr990125q}).

\bibitem{Chakrabarti2009}
Chakrabarti, D., Fejer, S.~N. \& Wales, D.~J., 2009 {Rational design of helical
  architectures}.
\newblock \emph{Proceedings of the National Academy of Sciences} \textbf{106},
  20164--20167.
\newblock ISSN 0027-8424.
\newblock (\doi{10.1073/pnas.0906676106}).

\bibitem{Douglas2009}
Douglas, J.~F., 2009 {Theoretical issues relating to thermally reversible
  gelation by supermolecular fiber formation}.
\newblock \emph{Langmuir} \textbf{25}, 8386--8391.
\newblock ISSN 07437463.
\newblock (\doi{10.1021/la9016245}).

\bibitem{Prybytak2012}
Prybytak, P., Frith, W.~J. \& Cleaver, D., 2012 {Hierarchical self-assembly of
  chiral fibres from achiral particles}.
\newblock \emph{Interface Focus} \textbf{2}, 651--657.
\newblock ISSN 2042-8898.

\bibitem{Wang2013}
Wang, Y., Xu, J., Wang, Y. \& Chen, H., 2013 {Emerging chirality in
  nanoscience.}
\newblock \emph{Chemical Society reviews} \textbf{42}, 2930--62.
\newblock ISSN 1460-4744.
\newblock (\doi{10.1039/c2cs35332f}).

\bibitem{Prockop1998}
Prockop, D.~J. \& Fertala, A., 1998 {The Collagen Fibril: The Almost
  Crystalline Structure}.
\newblock \emph{Journal of Structural Biology} \textbf{122}, 111--118.
\newblock ISSN 1047-8477.
\newblock (\doi{http://dx.doi.org/10.1006/jsbi.1998.3976}).

\bibitem{wess2008collagen}
Wess, T.~J., 2008 {Collagen fibrillar structure and hierarchies}.
\newblock In \emph{Collagen}, chapter~3, pp. 49--80. Springer.

\bibitem{Fratzl2003}
Fratzl, P., 2003 {Cellulose and collagen: from fibres to tissues}.
\newblock \emph{Current Opinion in Colloid {\&} Interface Science} \textbf{8},
  32--39.
\newblock ISSN 1359-0294.
\newblock (\doi{http://dx.doi.org/10.1016/S1359-0294(03)00011-6}).

\bibitem{Weisel1987}
Weisel, J.~W., Nagaswami, C. \& Makowski, L., 1987 {Twisting of fibrin fibers
  limits their radial growth.}
\newblock \emph{Proceedings of the National Academy of Sciences} \textbf{84},
  8991--8995.
\newblock ISSN 0027-8424.
\newblock (\doi{10.1073/pnas.84.24.8991}).

\bibitem{Rubin2008}
Rubin, N., Perugia, E., Goldschmidt, M., Fridkin, M. \& Addadi, L., 2008
  {Chirality of Amyloid Suprastructures}.
\newblock \emph{Journal of the American Chemical Society} \textbf{130},
  4602--4603.
\newblock ISSN 0002-7863.
\newblock (\doi{10.1021/ja800328y}).

\bibitem{Makowski1986}
Makowski, L. \& Magdoff-Fairchild, B., 1986 {Polymorphism of sickle cell
  hemoglobin aggregates: structural basis for limited radial growth}.
\newblock \emph{Science} \textbf{234}, 1228--1231.
\newblock ISSN 0036-8075.
\newblock (\doi{10.1126/science.3775381}).

\bibitem{Kleman1989}
Kl{\'{e}}man, M., 1989 {Curved crystals, defects and disorder}.
\newblock \emph{Advances in Physics} \textbf{38}, 605--667.
\newblock (\doi{10.1080/00018738900101152}).

\bibitem{Sadoc2006}
Sadoc, J.-F. \& Mosseri, R., 2006 \emph{{Geometrical Frustration}}.
\newblock Cambridge: Cambridge University Press.
\newblock ISBN 0 521 44198 6.

\bibitem{Grason2016}
Grason, G.~M., 2016 {Perspective: Geometrically frustrated assemblies}.
\newblock \emph{The Journal of Chemical Physics} \textbf{145}.
\newblock (\doi{http://dx.doi.org/10.1063/1.4962629}).

\bibitem{Grason2009}
Grason, G.~M., 2009 {Braided bundles and compact coils: The structure and
  thermodynamics of hexagonally packed chiral filament assemblies}.
\newblock \emph{Physical Review E} \textbf{79}, 041919.
\newblock ISSN 1539-3755.
\newblock (\doi{10.1103/PhysRevE.79.041919}).

\bibitem{Turner2003a}
Turner, M.~S., Briehl, R.~W., Ferrone, F.~A. \& Josephs, R., 2003 {Twisted
  protein aggregates and disease: the stability of sickle hemoglobin fibers.}
\newblock \emph{Physical review letters} \textbf{90}, 128103.
\newblock ISSN 0031-9007.
\newblock (\doi{10.1103/PhysRevLett.90.128103}).

\bibitem{Grason2007a}
Grason, G.~M. \& Bruinsma, R.~F., 2007 {Chirality and Equilibrium Biopolymer
  Bundles}.
\newblock \emph{Physical Review Letters} \textbf{99}, 098101.
\newblock ISSN 0031-9007.
\newblock (\doi{10.1103/PhysRevLett.99.098101}).

\bibitem{Yang2010}
Yang, Y., Meyer, R.~B. \& Hagan, M.~F., 2010 {Self-Limited Self-Assembly of
  Chiral Filaments}.
\newblock \emph{Physical Review Letters} \textbf{104}, 258102.
\newblock ISSN 0031-9007.
\newblock (\doi{10.1103/PhysRevLett.104.258102}).

\bibitem{Brown2014}
Brown, A.~I., Kreplak, L. \& Rutenberg, A.~D., 2014 {An equilibrium
  double-twist model for the radial structure of collagen fibrils}.
\newblock \emph{Soft Matter} \textbf{10}, 8500--8511.
\newblock ISSN 1744-683X.
\newblock (\doi{10.1039/C4SM01359J}).

\bibitem{Grason2015}
Grason, G.~M., 2015 {$\backslash$textit{\{}Colloquium{\}} : Geometry and
  optimal packing of twisted columns and filaments}.
\newblock \emph{Reviews of Modern Physics} \textbf{87}, 401--419.

\bibitem{Bruss2012a}
Bruss, I.~R. \& Grason, G.~M., 2012 {Non-Euclidean geometry of twisted filament
  bundle packing}.
\newblock \emph{Proceedings of the National Academy of Sciences} \textbf{109},
  10781--10786.
\newblock ISSN 0027-8424.
\newblock (\doi{10.1073/pnas.1205606109}).

\bibitem{Bruss2013}
Bruss, I.~R. \& Grason, G.~M., 2013 {Topological defects, surface geometry and
  cohesive energy of twisted filament bundles}.
\newblock \emph{Soft Matter} \textbf{9}, 8327--8345.
\newblock ISSN 1744-683X.
\newblock (\doi{10.1039/c3sm50672j}).

\bibitem{Bowick2009}
Bowick, M.~J. \& Giomi, L., 2009 {Two-dimensional matter: order, curvature and
  defects}.
\newblock \emph{Advances in Physics} \textbf{58}, 449--563.
\newblock ISSN 0001-8732.
\newblock (\doi{10.1080/00018730903043166}).

\bibitem{Manoharan2015}
Manoharan, V.~N., 2015 {Colloidal matter: Packing, geometry, and entropy}.
\newblock \emph{Science} \textbf{349}.

\bibitem{Seung1988}
Seung, H.~S. \& Nelson, D.~R., 1988 {Defects in flexible membranes with
  crystalline order}.
\newblock \emph{Physical Review A} \textbf{38}, 1005--1018.
\newblock ISSN 0556-2791.
\newblock (\doi{10.1103/PhysRevA.38.1005}).

\bibitem{Bausch2003a}
Bausch, A.~R., 2003 {Grain Boundary Scars and Spherical Crystallography}.
\newblock \emph{Science} \textbf{299}, 1716--1718.
\newblock ISSN 00368075.
\newblock (\doi{10.1126/science.1081160}).

\bibitem{Grason2010a}
Grason, G.~M., 2010 {Topological Defects in Twisted Bundles of
  Two-Dimensionally Ordered Filaments}.
\newblock \emph{Physical Review Letters} \textbf{105}, 045502.
\newblock ISSN 0031-9007.
\newblock (\doi{10.1103/PhysRevLett.105.045502}).

\bibitem{Grason2012}
Grason, G.~M., 2012 {Defects in crystalline packings of twisted filament
  bundles. I. Continuum theory of disclinations}.
\newblock \emph{Physical Review E} \textbf{85}, 031603.
\newblock ISSN 1539-3755.
\newblock (\doi{10.1103/PhysRevE.85.031603}).

\bibitem{Schneider2005}
Schneider, S. \& Gompper, G., 2005 {Shapes of crystalline domains on spherical
  fluid vesicles}.
\newblock \emph{Europhysics Letters (EPL)} \textbf{70}, 136--142.
\newblock ISSN 0295-5075.
\newblock (\doi{10.1209/epl/i2004-10464-2}).

\bibitem{Morozov2010}
Morozov, A.~Y. \& Bruinsma, R.~F., 2010 {Assembly of viral capsids, buckling,
  and the Asaro-Grinfeld-Tiller instability}.
\newblock \emph{Physical Review E} \textbf{81}, 041925.
\newblock ISSN 1539-3755.
\newblock (\doi{10.1103/PhysRevE.81.041925}).

\bibitem{Meng2014}
Meng, G., Paulose, J., Nelson, D.~R. \& Manoharan, V.~N., 2014 {Elastic
  instability of a crystal growing on a curved surface.}
\newblock \emph{Science (New York, N.Y.)} \textbf{343}, 634--7.
\newblock ISSN 1095-9203.
\newblock (\doi{10.1126/science.1244827}).

\bibitem{Hall2016}
Hall, D.~M., Bruss, I.~R., Barone, J.~R. \& Grason, G.~M., 2016 {Morphology
  selection via geometric frustration in chiral filament bundles}.
\newblock \emph{Nature Materials} p. 1614989.
\newblock ISSN 1476-1122.
\newblock (\doi{10.1038/nmat4598}).

\bibitem{Selinger1991}
Selinger, J.~V. \& Bruinsma, R.~F., 1991 {Hexagonal and nematic phases of
  chains. I. Correlation functions}.
\newblock \emph{Phys. Rev. A} \textbf{43}, 2910--2921.
\newblock (\doi{10.1103/PhysRevA.43.2910}).

\bibitem{deGennes1995physics}
de~Gennes, P.~G. \& Prost, J., 1995 \emph{{The physics of liquid crystals}}.
\newblock 83. Oxford university press.
\newblock ISBN 0198517858, 9780198517856.

\bibitem{Kamien1996}
Kamien, R.~D. \& Nelson, D.~R., 1996 {Defects in chiral columnar phases:
  Tilt-grain boundaries and iterated moir{\'{e}} maps}.
\newblock \emph{Phys. Rev. E} \textbf{53}, 650--666.
\newblock (\doi{10.1103/PhysRevE.53.650}).

\bibitem{Wright1989a}
Wright, D.~C. \& Mermin, N.~D., 1989 {Crystalline liquids: The blue phases}.
\newblock \emph{Reviews of Modern Physics} \textbf{61}, 385--432.
\newblock ISSN 00346861.
\newblock (\doi{10.1103/RevModPhys.61.385}).

\bibitem{Azadi2016}
Azadi, A. \& Grason, G.~M., 2016 {Neutral versus charged defect patterns in
  curved crystals}.
\newblock \emph{Physical Review E} \textbf{94}, 013003.
\newblock ISSN 2470-0045.
\newblock (\doi{10.1103/PhysRevE.94.013003}).

\bibitem{Kohler2016}
K{\"{o}}hler, C., Backofen, R. \& Voigt, A., 2016 {Stress Induced Branching of
  Growing Crystals on Curved Surfaces}.
\newblock \emph{Physical Review Letters} \textbf{116}, 135502.
\newblock ISSN 0031-9007.
\newblock (\doi{10.1103/PhysRevLett.116.135502}).

\end{thebibliography}
\end{document}